\documentclass[aps,prb,showpacs,notitlepage,floatfix,twocolumn,superscriptaddress,eqsecnum]{revtex4-2}
\usepackage[bookmarks=false,colorlinks=true,urlcolor=red3,citecolor=red3,linkcolor=red3]{hyperref}
\usepackage{graphicx,amsmath,amssymb,amsxtra,comment}
\usepackage{wasysym}
\usepackage{xcolor,dsfont}
\definecolor{magenta}{rgb}{1,0,1}
\usepackage{tikz,rotating}
\usetikzlibrary{arrows.meta}
\usepackage{ellipse}
\usepackage{booktabs} 

\def\bd{\begin{displaymath}}
\def\be{\begin{equation}}
\def\ed{\end{displaymath}}
\def\ee{\end{equation}}
\def\bsub{\begin{subequations}}
\def\esub{\end{subequations}}

\usepackage{tikz}
\usetikzlibrary{shapes}
\makeatletter
\newcommand*\dashline{\rotatebox[origin=c]{90}{$\dabar@\dabar@\dabar@$}}
\makeatother

\def\bra#1{\left\langle#1\right|}
\def\ket#1{\left|#1\right\rangle}

\def\be{\begin{equation}}
\def\ee{\end{equation}}

\def\bea{\begin{eqnarray}}
\def\eea{\end{eqnarray}}

\usepackage{subfigure}
\usepackage{times}
\usepackage{graphicx}
\usepackage{xcolor,dsfont}

\graphicspath{{figures}}

\usepackage{autobreak}
\allowdisplaybreaks

\RequirePackage{color}

\definecolor{MyDarkGreen}{rgb}{0.02,0.60,0.06}
\definecolor{darkgreen}{rgb}{0.02,0.60,0.06}

\definecolor{kellygreen}{rgb}{0.3, 0.73, 0.09}
\definecolor{deeppink}{rgb}{1.0, 0.08, 0.58}
\definecolor{myblue}{HTML}{0096FF}

\definecolor{red3}{HTML}{F68B1F} 
\definecolor{red3}{rgb}{0.3, 0.73, 0.09} 
\definecolor{red3}{HTML}{C04DFF} 

\graphicspath{{figures}}

\begin{document}
\title{Exactly solvable non-planar $\mathbb{Z}_{2}$ dimer liquids on checkerboard and ruby lattices}

\author{Julia Wildeboer}
\email{jwildeboe@bnl.gov}
\affiliation{Division of Condensed Matter Physics and Materials Science,
Brookhaven National Laboratory, Upton, NY 11973-5000, USA}
\affiliation{Perimeter Institute for Theoretical Physics, Waterloo, Ontario, N2L 2Y5, Canada}

\author{Zohar Nussinov}
\affiliation{Department of Physics, Washington University, St. Louis, Missouri 63130, USA}
\affiliation{Institut für Physik, Technische Universität Chemnitz, 09107 Chemnitz, Germany}
\affiliation{LPTMC, CNRS-UMR 7600, Sorbonne Universit\'e, 4 Place Jussieu, 75252 Paris cedex 05, France}

\author{Thomas Iadecola}
\affiliation{Department of Physics, The Pennsylvania State University, University Park, Pennsylvania 16802, USA}
\affiliation{Institute for Computational and Data Sciences, The Pennsylvania State University, University Park, Pennsylvania 16802, USA}
\affiliation{Materials Research Institute, The Pennsylvania State University, University Park, Pennsylvania 16802, USA}

\author{Alexander Seidel}
\affiliation{Department of Physics, Washington University, St. Louis, Missouri 63130, USA}

\begin{abstract}
We generalize an influential framework for exactly solvable quantum
dimer models with dual Ising gauge theory descriptions realizing the
$\mathbb{Z}_{2}$ topological phase from trivalent to tetravalent parent
lattices. The resulting quantum dimer models live on crossed-medial
lattices and possess the remarkable feature of having crossed
plaquettes, leading to transition graphs with generically intersecting
loops. This non-planar structure runs counter to established templates
for analytically tractable dimer liquids.
As the simplest realization of the construction, we introduce an
exactly solvable quantum dimer model on the checkerboard lattice that allows an exact mapping to the toric code, thus providing a particularly direct connection between the latter and Anderson's short range resonating valence bond paradigm.
We further show that a corresponding crossed ruby lattice construction, dual to an Ising gauge theory on the kagome lattice, naturally falls within the same framework. 
More generally, the construction gives rise to a broad class of exactly
solvable crossed-medial quantum dimer models and admits natural
iteration, generating cascades of solvable $\mathbb{Z}_{2}$ topological
dimer liquids beyond the standard planar setting. We furthermore extend Kasteleyn methods to the relevant non-planar graphs, enabling controlled wave-function deformations away from the commuting-projector points while retaining efficient evaluation of correlation functions. 
\end{abstract}

\maketitle
\tableofcontents
\section{Introduction}\label{section_intro}
Quantum dimer models (QDMs) provide a particularly useful framework for studying strongly correlated phases in which local constraints, fractionalized excitations, and topological order play a central role. Originally introduced by Rokhsar and Kivelson~\cite{RK}, these models capture the essential physics of Anderson's short-ranged resonating-valence-bond (RVB) states~\cite{anderson1987resonating,fazekas1974ground} while remaining far more tractable than the underlying frustrated spin systems from which they may emerge. In particular, QDMs have served as paradigmatic settings for realizing quantum dimer liquids, deconfined excitations, and topological order~\cite{wen1990topological, MS,misguich,ioffe2002topologically, kitaev2003fault, moessner2011quantum}.

A major reason for this success is that certain dimer models admit exact treatments. On the classical side, Kasteleyn showed that close-packed dimers on planar graphs can be counted by Pfaffian methods~\cite{kas}, a result further developed by Fisher~\cite{Fisher} and by Fisher and Stephenson~\cite{FS}. On the quantum side, exactly solvable Rokhsar--Kivelson points furnish, among other things, controlled examples of topological liquids~\cite{MS}. 
More generally, QDMs provide a bridge between frustrated magnetism, constrained many-body dynamics, and lattice gauge theory~\cite{Fradkin1990,Moessner2001, misguich2003, MS2003, Ardonne2004, Henley2004, Nogueira2009, FradkinBook, ShahPRX}. They have also recently become relevant in the broader context of constrained quantum dynamics and quantum many-body scar phenomena~\cite{wildeboer2020topological}.

Most analytically controlled dimer constructions have focused on planar lattices. As we showed in our earlier work on the classical problem~\cite{wildeboer2020checkerboard}, however, planarity is not a fundamental requirement. This observation motivates the search for quantum dimer models on non-planar lattices that remain exactly solvable and continue to exhibit topological order. The present work provides such a construction.

A particularly influential class of fully solvable QDMs arises from
the construction of Misguich, Serban, and Pasquier~\cite{misguich},
which exactly maps quantum dimer liquids  on certain lattices, notably the kagome,
to $\mathbb{Z}_{2}$ lattice gauge theories on trivalent parent lattices.
 The QDMs in question live on medial lattices and admit exact
commuting-projector formulations with $\mathbb{Z}_{2}$ topological order. In the present work, we generalize this framework from trivalent to tetravalent parent lattices. 
As a new feature, the resulting medial lattices must be adorned with crossing diagonals on rectangular plaquettes, together with special local constraints, leading to non-planar dimer configurations whose transition graphs generically contain intersecting loops. This crossed-medial structure runs counter to established templates for analytically controlled dimer liquids and calls for generalization of several standard tools, including Kasteleyn methods.

An advantage of our construction is the fact that the square lattice now becomes an admissible lattice for hosting the $\mathbb{Z}_{2}$ gauge theory formulation of the model. The corresponding QDM then lives on the checkerboard lattice -- a square lattice with every second plaquette decorated via diagonal links. This then leads not only to the simplest instance of our model, but one that can be directly mapped, via square lattice Ising gauge theory (IGT), to the original toric code~\cite{kitaev2003fault}.
An appealing feature of the checkerboard realization is thus that it provides a particularly direct connection between two historically important viewpoints on the $\mathbb{Z}_2$
topological phase: Anderson's short-range RVB paradigm of quantum dimer liquids~\cite{Anderson1973} and the toric code, now widely regarded as a canonical fixed-point Hamiltonian for this phase.
Using this simplest realization of
the framework, we analyze the resulting phase in detail, exposing it as a fully gapped $\mathbb{Z}_{2}$ dimer liquid with ultra-local equal-time correlations, deconfined vison excitations, topological ground-state degeneracy, and an exact IGT dual description. We also analyze its entanglement structure and modular data, which are consistent with toric-code topological order.

Another attractive lattice where our  framework leads to a new commuting-projector QDM with $\mathbb{Z}_{2}$ topological ground state
is the ruby lattice, i.e., the medial lattice of the kagome lattice. The ruby lattice has recently been recognized as a particularly rich
arena for constrained dynamics and emergent $\mathbb{Z}_{2}$ topological
order~\cite{Verresen2021,Verresen2022}. Our work provides an exactly
solvable realization of this physics within a generalized crossed-medial
dimer framework. Unlike the square lattice, the ruby lattice is not self-medial. As any medial lattice of a two-dimensional lattice is tetravalent, any such instance of our model can be seen as a member of an infinite cascade of models within the paradigm established here. On the other hand, while the square lattice does not give rise to such an infinite cascade, its property of being self-medial is responsible for the direct equivalence between our checkerboard model and the toric code.

Using mostly the ruby lattice version of our framework as an example, we will also briefly elaborate on wave function deformations of our models that interpolate between planar and non-planar cases, and, despite including the integrable non-planar points, can be treated efficiently via Kasteleyn methods.

The remainder of this paper is organized as follows. In Secs.~\ref{section_constraints} and \ref{section_arrow_formalism}, we introduce the checkerboard lattice version of the model and the associated pseudospin (or ``arrow") representation~\cite{elser1993kagome}. In Secs.~\ref{section_arrow_ham} and \ref{section_QDM_model}, we construct the exactly solvable arrow and dimer Hamiltonians. Section~\ref{section_corrgap} discusses ultra-local
correlations and localized vison excitations, while Sec.~\ref{section_Ising} develops the exact mapping to $\mathbb{Z}_{2}$ gauge theory. Section~\ref{section_ee} addresses entanglement and minimum-entropy states. In Sec.~\ref{section_ruby}, we formulate the ruby-lattice version as well as further generalizations of the model, and contrast the regular and crossed ruby cases, giving the nuts and bolts of a Kasteleyn-like formalism that incorporates both.
Two Appendices develop the Kasteleyn formalism in more detail and include checks on the non-planar versions.

\section{Structural constraints and dimerization patterns in the checkerboard dimer model}\label{section_constraints}
\begin{figure}[b]
\includegraphics[width=1.0\columnwidth]{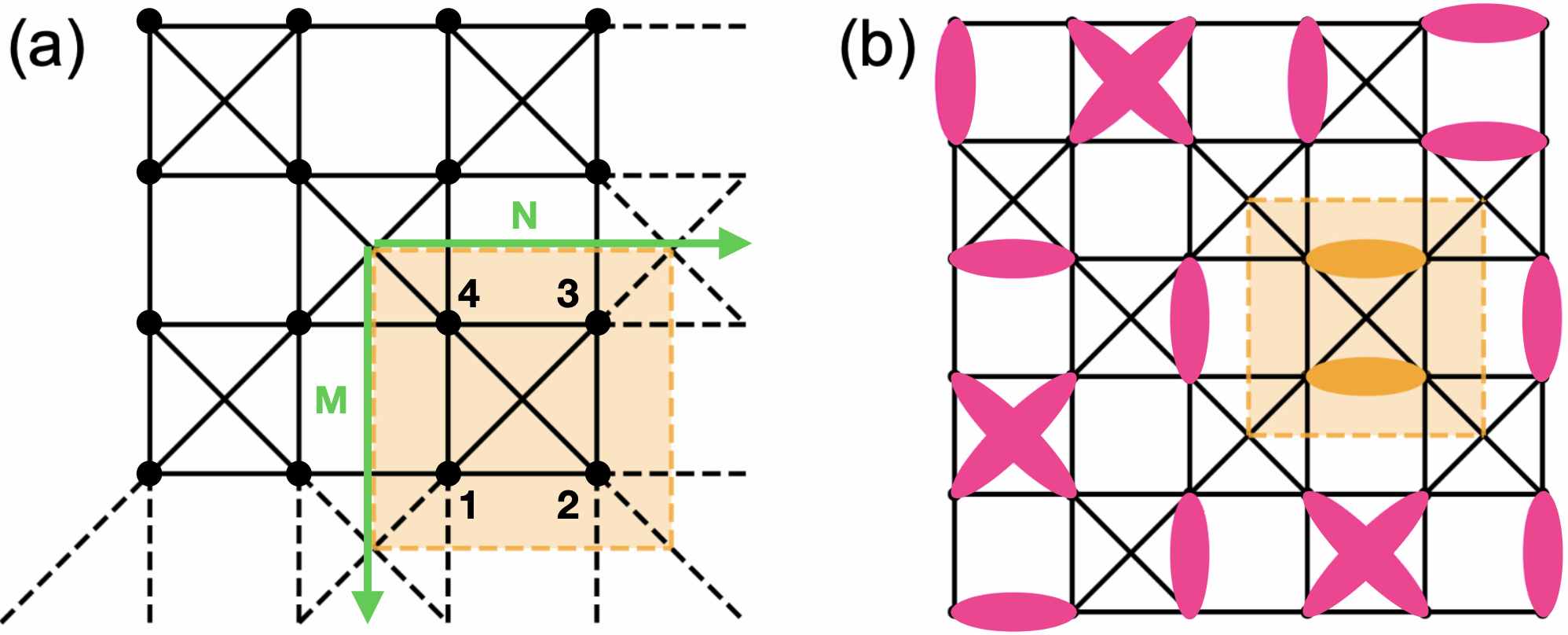}
\caption{
        {\it Checkerboard lattice geometry and local crossed-plaquette constraint.} 
        (a) Checkerboard lattice with the four-site unit cell shaded. The sites within the unit cell are labeled $1,2,3,4$, and the two lattice directions are indexed by $M$ and $N$. 
        (b) Forbidden local dimer configuration on a crossed plaquette: two parallel dimers may not occupy the same crossed plaquette. Whenever two dimers occupy a crossed plaquette, they must instead form a crossed pair on the two diagonal links. This local restriction defines the constrained Hilbert space of the model.
        }
\label{figure_666}
\end{figure}

We begin by specifying the lattice geometry and the local constraint that defines the checkerboard dimer model. Figure~\ref{figure_666}(a) shows the four-site unit cell, which serves as the basic building block of the lattice. As in a conventional close-packed dimer model, we impose the hard-core constraint that every lattice site is touched by exactly one dimer. 

In addition to this standard constraint, our model excludes a specific class of local configurations on crossed plaquettes. As illustrated in Fig.~\ref{figure_666}(b), a crossed plaquette may not be occupied by two {\it parallel} dimers. Equivalently, whenever two dimers occupy the same crossed plaquette, they must form a crossed pair on the two diagonal links. This local restriction is the defining structural ingredient of the model.

The constrained Hilbert space that results from this rule is
considerably smaller than that of the unconstrained checkerboard dimer problem and, as we show below, admits a simple arrow representation that makes the exact solution possible. An example of an allowed dimer covering is shown in Fig.~\ref{figure1}(a), and the corresponding arrow formulation is developed in Sec.~\ref{section_arrow_formalism}. 

\section{Arrow formalism in the checkerboard dimer model}\label{section_arrow_formalism}
\begin{figure}[t]
\includegraphics[width=1.00\columnwidth]{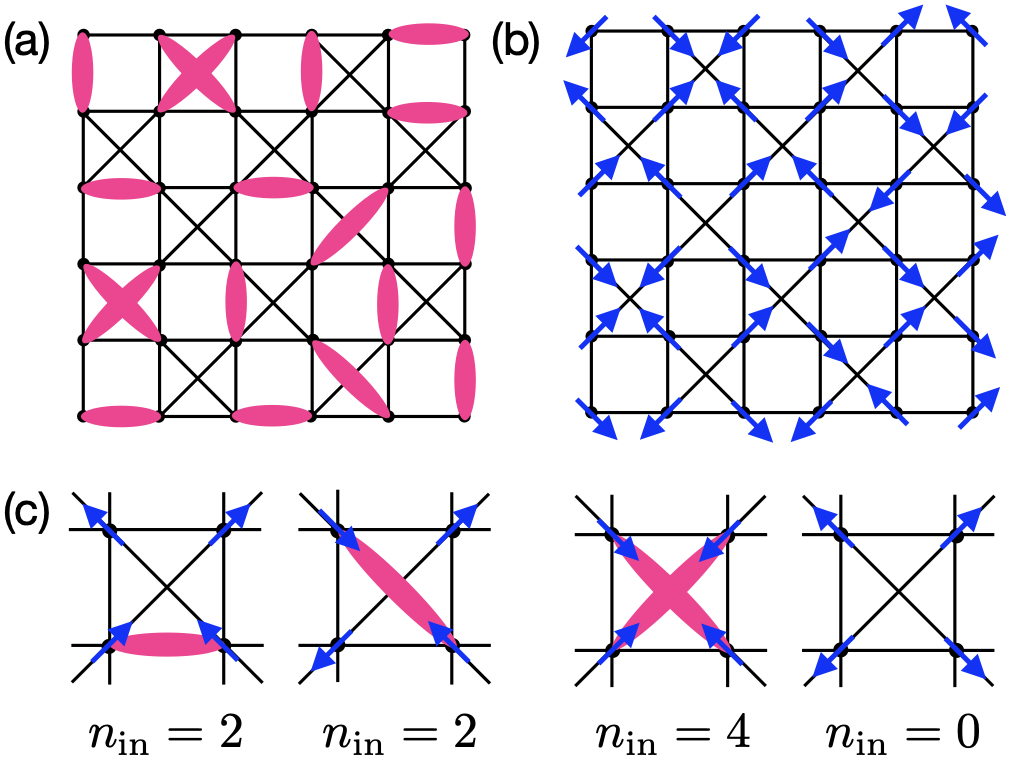}
\caption{   
      {\it Arrow representation of admissible checkerboard dimer coverings.}
      (a) Example of an allowed dimer covering of the checkerboard model. Whenever a crossed plaquette is occupied by two dimers, they form a crossed pair on the diagonal links. 
      (b) The same configuration represented in terms of arrow variables. 
      (c) Allowed local arrow configurations on a crossed plaquette. The number of arrows pointing toward the plaquette center must be even, i.e. $n_{\mathrm{in}}=0,2,4$. The two cases with $n_{\mathrm{in}}=2$ correspond to plaquettes occupied by a single dimer, the case $n_{\mathrm{in}}=4$ corresponds to a crossed pair of dimers on the two diagonal links, and the case $n_{\mathrm{in}}=0$ corresponds to an empty crossed plaquette.    
}
\label{figure1}
\end{figure}

Elser and Zeng's seminal work~\cite{elser1993kagome} established a correspondence between dimer coverings on the kagome lattice links and configurations of arrows positioned on the lattice vertices, enabling an alternative representation of dimer arrangements and associated constraints. Interestingly, this formalism can be extended to the checkerboard dimer model in the presence of the constraints discussed above [Figures~ ~\ref{figure_666}(b) and \ref{figure1}(a)]. To see this, we associate an arrow with each site of the checkerboard lattice.
Since every site belongs to two neighboring crossed plaquettes, the arrow may point toward the center of either one of these two plaquettes [see Fig.~\ref{figure1}(b,c)]. 
Additionally, we enforce the constraint that the number of inward pointing arrows at each crossed plaquette must be {\em even}, i.e., 
\begin{equation}
n_{\mathrm{in}}=0,2,4\,.
\end{equation}
Link occupancies can now be resolved for every crossed plaquette as follows: If all arrows on the plaquette point outward, the links of the plaquette do not carry dimers. If two arrows on the plaquette point inward, the link connecting the sites of these arrows (horizontal, vertical, or diagonal) will then be the only link of the plaquette that is occupied by a dimer. If all four arrows of the plaquette point inward, the crossed diagonal links are both occupied by a dimer (consequently, the horizontal and vertical links are empty). These rules facilitate a one-to-one correspondence between arrow configurations and dimer configurations, subject to the respective constraints.
\begin{figure*}[t]
\includegraphics[width=1.0\textwidth]{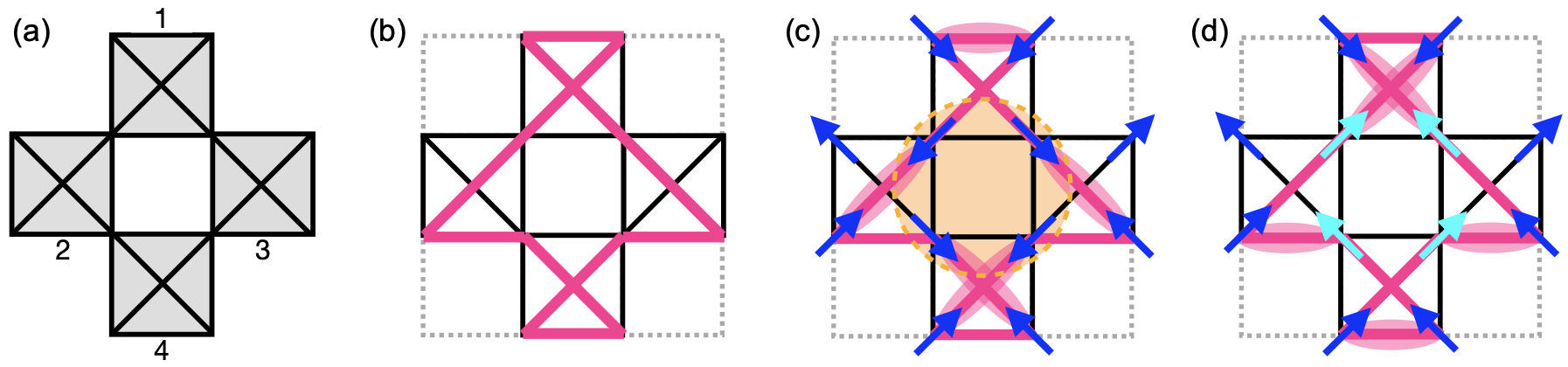}
\caption{    
        {\it Local resonance move on a checkerboard superplaquette.}
        (a) A 12-site superplaquette consisting of a central uncrossed plaquette (white) surrounded by four crossed plaquettes (gray), labeled $1,\ldots,4$.  
        (b) A representative resonance loop on the superplaquette. 
        (c) Allowed arrow configuration together with the associated resonance loop and one dimerization pattern on that loop. 
        (d) After flipping the four arrows on the central uncrossed plaquette [highlighted in orange in (c) and shown in cyan in (d)], the resonance loop itself is unchanged, while the dimers shift to the complementary set of links along the same loop. This illustrates how the local arrow flip becomes a local dimer resonance move.
        }
\label{figure16}
\end{figure*}

We consider systems with periodic boundary conditions containing an even number of sites (having an equal number of crossed and uncrossed plaquettes). If the lattice contains a total number $N_{s}$ of sites, then it also contains $N_{s}$ arrow variables and $N_{s}/2$ crossed plaquettes. Each crossed plaquette imposes one parity constraint. However, these $N_{s}/2$ constraints are not all independent. Since every site belongs to two crossed plaquettes, the product of all crossed-plaquette parity constraints is automatically equal to unity. Thus only $\frac{N_{s}}{2}-1$ 
of the local constraints are independent. 
This leaves exactly 
\begin{eqnarray}
\mathcal{F}_{\rm Ising} = \frac{N_{s}}{2}+1    
\end{eqnarray}
independent Ising degrees of freedom, which define the configuration space of the system. Consequently, the total number of permissible dimer coverings is  
\begin{eqnarray}
\mathcal{N}_{\rm dimer} =2^{N_s/2+1}\,.
\end{eqnarray}
This counting is directly analogous to the kagome-lattice hard-core dimer problem~\cite{elser1993kagome,phares1988thermodynamics,elser1989nuclear}. Within a fixed topological sector (see Sec.~\ref{section_QDM_model} and following sections), two additional global constraints apply, reducing the number of dimers to 
$\mathcal{N}_{\mathrm{dimer}}/4=2^{N_{s}/2-1}$.
The arrow representation not only provides an alternative description
of the constrained dimer Hilbert space, but also supplies the
structural bridge between the quantum dimer model and its dual
$\mathbb{Z}_2$ lattice gauge theory formulation, which ultimately
underlies the exact solvability of the model constructed in Sections~\ref{section_arrow_ham} and \ref{section_QDM_model}.

\section{Arrow representation and exactly solvable Hamiltonian}\label{section_arrow_ham}
In the following two sections, we construct an exactly solvable Hamiltonian whose ground states are equal-amplitude superpositions of all valid dimer coverings.
More precisely, our goal is to realize ground states of the Rokhsar--Kivelson form 
\begin{eqnarray}\label{wf}
\ket{\Psi} = \frac{1}{{\sqrt{\mathcal{N}}}}\sum_{D} \ket{D}\,. 
\end{eqnarray}
The sum runs over all valid dimer coverings $\ket{D}$ (possibly restricted to a topological sector) and $\mathcal{N}$ denotes their total number. 
Here, the validity of a dimer covering in particular
requires that the crossed-plaquette constraint 
described above is satisfied, namely that parallel dimers on crossed plaquettes are forbidden (see, e.g., Fig.~\ref{figure_666}(b) and caption).
\begin{figure}[t]
\includegraphics[width=1.00\columnwidth]{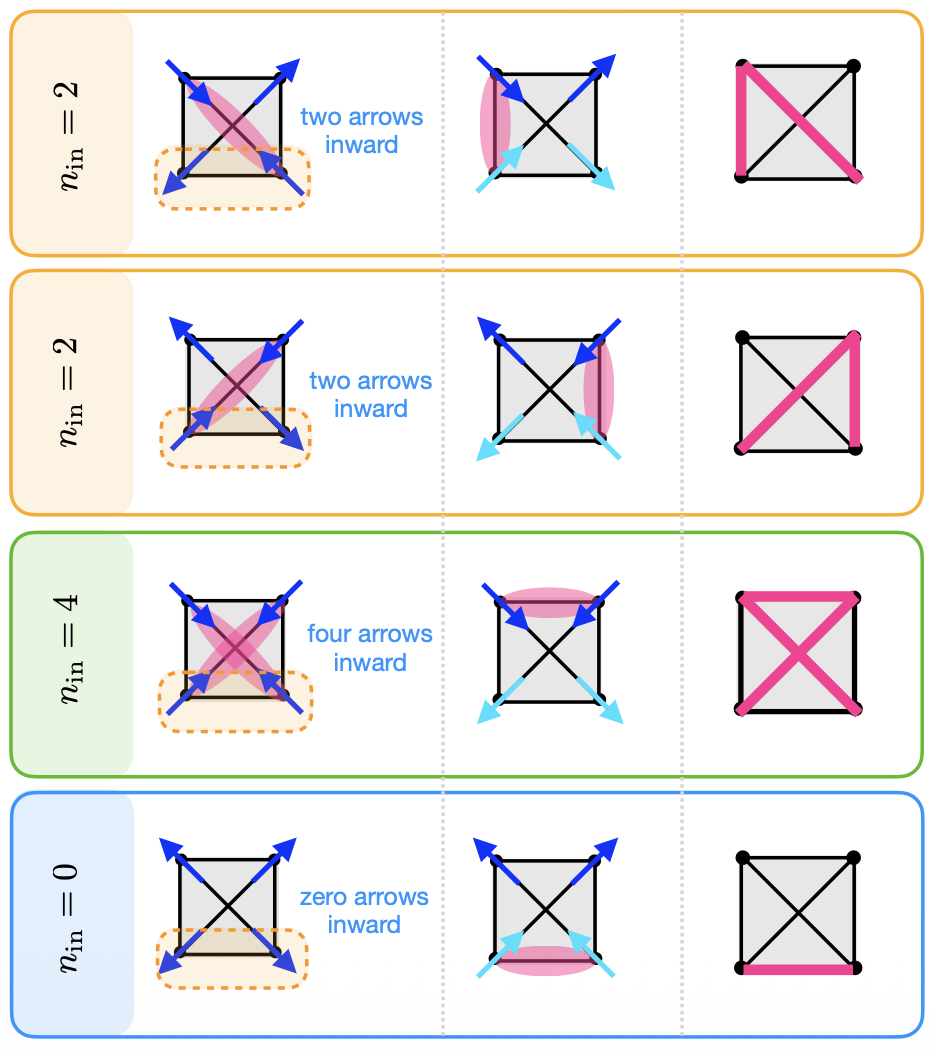}
\caption{
        {\it Local rule for constructing resonance-loop segments from arrow flips.}
        For a resonance loop associated with a central uncrossed plaquette, the figure illustrates the local construction rule for the crossed plaquette located directly above the central plaquette. 
        Left column: allowed arrow configuration on that crossed plaquette before flipping the central uncrossed plaquette. Middle column: configuration obtained after flipping the two arrows shared with the central uncrossed plaquette.
        Right column: the resulting local loop segment passing through the crossed plaquette. The four rows correspond to the distinct local cases entering the loop construction: two inequivalent $n_{\mathrm{in}}=2$ configurations, the $n_{\mathrm{in}}=4$ crossed-pair configuration, and the $n_{\mathrm{in}}=0$ empty-plaquette configuration. The analogous rules for the crossed plaquettes to the left, right, and below follow by symmetry, and assembling the four local segments produces the closed resonance loop shown in Fig.~\ref{figure16}.        
        }
\label{figure4}
\end{figure}

At this stage, it is convenient to postpone the formulation of the Hamiltonian in the dimer basis and work instead in the arrow representation. 
Similar to the kagome lattice QDM, we may utilize this representation directly to construct a set of commuting local operators. More specifically, for each {\it uncrossed} plaquette $\square$, we define
\begin{align}
\begin{split}
\sigma^{x}(\square)&=\prod_{i\,\in\, \square}\sigma_i^x \\
&=
\sigma_1^x(\square)\sigma_2^x(\square)
\sigma_3^x(\square)\sigma_4^x(\square)
\label{operator1}
\end{split}
\end{align}
where the four operators $\sigma_{i}^{x}$ flip the arrows on the four vertices of the plaquette $\square$. Equivalently, $\sigma^{x}(\square)$  may be written as a sum over all local arrow-flip processes on that plaquette $\square$, i.e. 
\begin{equation}
\sigma^{x}(\square)
=
\frac12\sum_{i=1}^{16}
\Big(
\ket{A_i(\square)}\bra{\overline A_i(\square)}
+
\ket{\overline A_i(\square)}\bra{A_i(\square)}
\Big)\,.
\label{operator1a}
\end{equation}
Here $\ket{A_{i}(\square)}$ and $\ket{\overline A_{i}(\square)}$ denote the two local arrow configurations related by reversing all four arrows on the uncrossed plaquette $\square$. Since the uncrossed plaquettes are unconstrained,
there are $2^4=16$ local arrow configurations in total, grouped into eight such pairs. 

From both \eqref{operator1} and \eqref{operator1a}, it is clear that $\sigma^{x}(\square)$ maps local configurations pairwise onto one another and so is involutive, i.e. 
$(\sigma^{x}(\square))^{2}= \mathds{1}$. 
We furthermore note that the action of
$\sigma^{x}(\square)$ clearly conserves the arrow constraint for all  crossed plaquettes, as it flips exactly two arrows on each adjacent crossed plaquette. Moreover, it is evident from Eq.~\eqref{operator1} that the operators $\sigma^{x}(\square)$ commute, even for pairs of corner-sharing uncrossed plaquettes. 
Equipped with these commuting operators, we now define the following arrow 
Hamiltonian, 
\begin{eqnarray}\label{hamil}
{\cal H} = -\gamma \sum_{\square}\sigma^{x}(\square)\,.
\end{eqnarray}

Since each $\sigma^{x}(\square)$ is an involutive operator satisfying 
$[\sigma^{x}(\square), \sigma^{x}(\square')] = 0$ for all plaquettes $\square$, $\square^{\prime}$, 
the full Hamiltonian is exactly solvable.  
In particular, all $\sigma^x(\square)$ can be diagonalized simultaneously.

Because $\sigma^{x}(\square)$ has eigenvalues $\lambda_{\square}=\pm 1$,
every eigenstate of the Hamiltonian corresponds to an energy 
\begin{equation}
E = -\gamma \sum_{\square} \lambda_{\square}\,.
\end{equation}
In the ground state sector, $\lambda_{\square}=+1$ for all uncrossed plaquettes $\square$.

A state $|\Psi\rangle$ is therefore a ground state of ${\cal H}$ if and only if it satisfies
\begin{equation}
\sigma^{x}(\square)\,|\Psi\rangle = |\Psi\rangle 
\quad \forall \text{\,\, uncrossed plaquettes }\square\,.
\label{eq:arrow_ground state_condition}
\end{equation}
In the presence of periodic boundary conditions, the arrow-constraint subspace decomposes into four topological sectors distinguished by
nonlocal $\mathbb{Z}_{2}$ invariants, which we will define more carefully in the following section. In particular, the dynamics of the Hamiltonian~\eqref{hamil} leave these sectors invariant and are, moreover, ergodic within each sector, as follows from general arguments presented in Sec.~\ref{rubyA}.
It follows that within each such sector $A$, there is a unique state obeying Eq.~\eqref{eq:arrow_ground state_condition}, namely the equal-weight superposition
\begin{equation}
|\Psi_{0}^{A}\rangle \;=\;
\frac{1}{\sqrt{|A|}}\sum_{\vec a\in A} |\vec a\rangle\, ,
\label{eq:arrow_equal_weight_state}
\end{equation}
where $|\vec a\rangle$ ranges over all arrow configurations in the sector $A$ that satisfy the local arrow constraints.
Each $|\Psi_{0}^{A}\rangle$ is a $+1$ eigenstate of every $\sigma^{x}(\square)$, and hence
\begin{equation}
{\cal H}\,|\Psi_{0}^{A}\rangle = 
-\gamma\, N_{\square}\, |\Psi_{0}^{A}\rangle\, ,
\end{equation}
with $N_{\square}$ denoting the number of uncrossed plaquettes in the system.

\section{Quantum Dimer Representation of the Arrow Hamiltonian}\label{section_QDM_model}
We now rewrite the arrow Hamiltonian Eq.~\eqref{hamil} in the standard orthonormal dimer basis of the quantum dimer model via the arrow $\longleftrightarrow$ dimer mapping introduced for the checkerboard lattice in Sec.~\ref{section_arrow_formalism}. By definition, this basis consists only of dimer coverings satisfying the crossed-plaquette constraint introduced in Sec.~\ref{section_constraints}.

It is clear that the arrow Hamiltonian defined in the preceding section is purely dynamic, and will consist of only non-diagonal ``resonance moves'' also when expressed in the dimer language.
The appropriate local object to host these moves is the superplaquette shown in Fig.~\ref{figure16}(a), consisting of a central uncrossed plaquette together with the four surrounding crossed plaquettes. On each of these four crossed plaquettes, the arrow configuration determines the local dimer occupancy uniquely, and hence fixes all links of the superplaquette. The resulting dimer configuration, or equivalently the corresponding arrow configuration, defines a closed loop of links surrounding the central plaquette. This loop construction is illustrated in Figs.~\ref{figure16} and \ref{figure4}. 
More specifically, each crossed plaquette contributes a local loop segment determined by the pair of arrow configurations related by the action of $\sigma^{x}(\square)$ on the central uncrossed plaquette. Figure~\ref{figure4} shows the distinct local cases and their associated loop segments. These local segments join together to form a single closed loop $C$ on the superplaquette, as shown in
Fig.~\ref{figure16}. Along this loop, the dimers occupy every second link. The action of $\sigma^{x}(\square)$ then flips the four arrows on the central uncrossed plaquette and thereby exchanges occupied and unoccupied links along the same loop. In this way, each local arrow flip becomes a local dimer resonance move.

\begin{figure*}[t]
\includegraphics[width=1.0\textwidth]{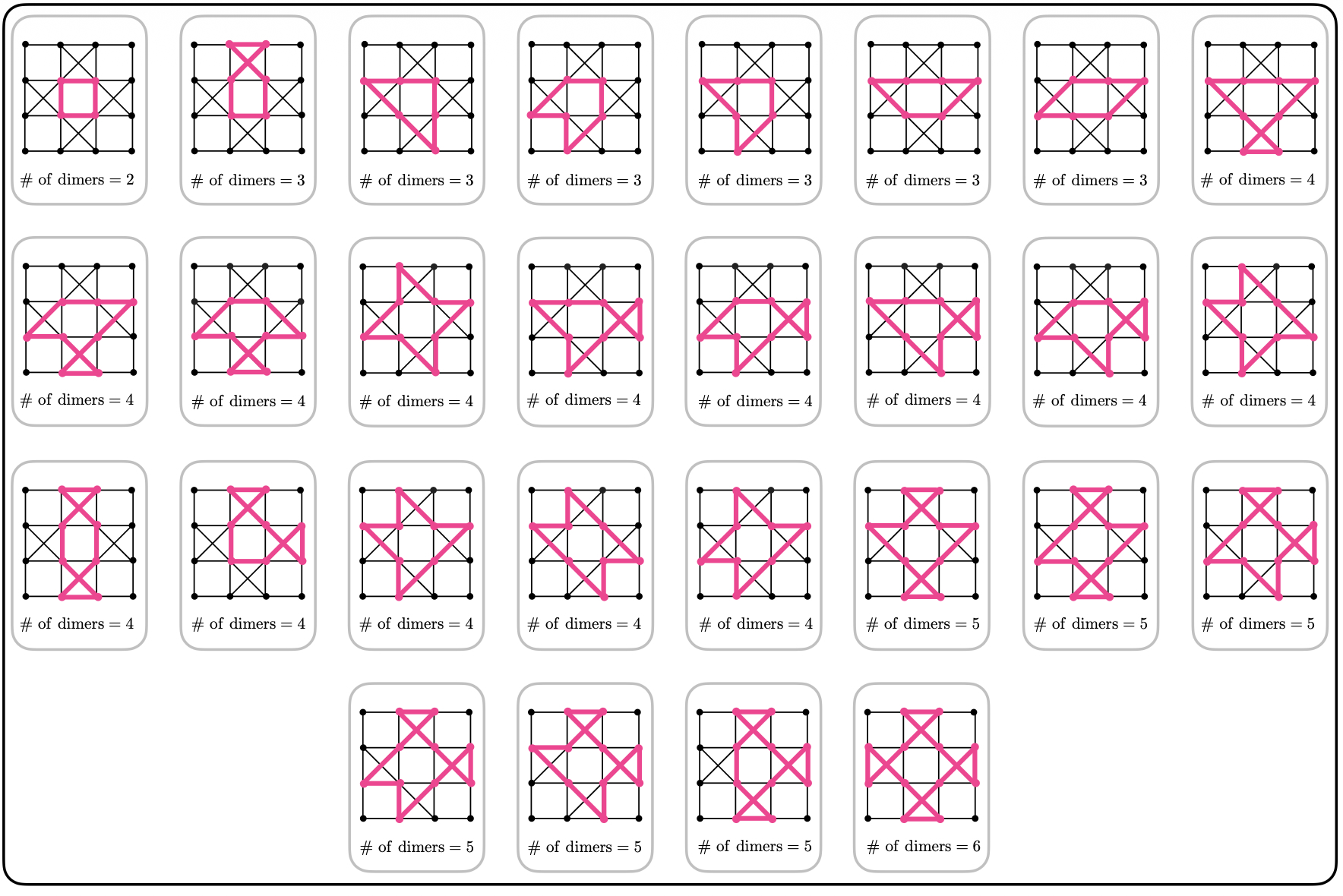}
\caption{    
         {\it Representative local resonance loops of the checkerboard dimer Hamiltonian.}
         One representative loop from each symmetry class of local resonance moves supported on a checkerboard superplaquette. The number of dimers participating in each resonance move is indicated below the corresponding diagram. This equals half of the loop length. Additional loops related by lattice symmetries are not shown.
}
\label{figure_loops}
\end{figure*}

The nonvanishing kinetic matrix elements of the dimer Hamiltonian are therefore in one-to-one correspondence with the closed loops that can occur on a superplaquette. We denote by $\mathcal{C}(\square)$ the set of all such closed loops supported on the superplaquette associated with the central uncrossed plaquette $\square$. For convenience,
Figure~\ref{figure_loops} lists one representative loop from each symmetry class. All other members of $\mathcal C(\square)$ are obtained from these representatives by lattice rotations and reflections. 

For each loop $C \in \mathcal C(\square)$, let $\ket{C}$ and
$\ket{\overline C}$ denote the two dimer configurations that differ only by exchanging occupied and unoccupied links along the loop $C$. The corresponding local resonance operator is
\begin{equation}
T_C = \ket{C}\bra{\overline C} + \ket{\overline C}\bra{C}\,.
\end{equation}
This is the dimer-basis counterpart of the local arrow-flip processes appearing in Eq.~\eqref{operator1a}. There are, however, many more loops $C$ than the eight local arrow pairs entering Eq.~\eqref{operator1a}, because the loop type depends on the arrow configuration on all sites of the superplaquette, not only on the four sites of the central uncrossed plaquette.

The full dimer Hamiltonian is obtained by summing these local resonance operators over all superplaquettes of the lattice, i.e. 
\begin{equation}
\mathcal{H}^{D} = -\gamma\sum_{\square}\sum_{C\in\mathcal C(\square)} T_C\,.
\label{eq:dimerHamiltonian_final}
\end{equation}
By construction, the matrix elements of $\mathcal{H}^D$ coincide exactly with those of the arrow Hamiltonian $\mathcal{H}$ in Eq.~\eqref{hamil}. In particular, for each uncrossed plaquette $\square$, the action of $\sigma^{x}(\square)$ is represented in the dimer basis by the sum of all resonance moves supported on the corresponding superplaquette 
\begin{equation}
\sigma^x(\square) \equiv \sum_{C\in\mathcal C(\square)} T_C\,.
\label{duality}
\end{equation}
Hence $\mathcal{H}^{D}$ provides a faithful dimer representation of the arrow model.

Since the matrix elements of ${\cal H}^{D}$ reproduce exactly those of the arrow Hamiltonian ${\cal H}$ of Eq.~\eqref{hamil}, the two models have identical spectra. In particular, their ground states coincide.  Expressed in the dimer basis, the ground state in each topological sector takes the Rokhsar--Kivelson form
\begin{equation}
|\Psi_{0}^{A}\rangle
= \frac{1}{\sqrt{|A|}} \sum_{D \in A} |D\rangle\, ,
\label{eq:dimer_GS}
\end{equation}
where $A$ labels a topological sector and the sum runs over all valid dimer coverings in that sector. Distinct sectors are characterized by the parity of dimers intersecting noncontractible loops on the lattice. That is, one may define winding operators $\Omega_{x}$ and $\Omega_{y}$ measuring the dimer occupation mod~2 along two fixed noncontractible cycles, one for each of the two lattice directions; the ground
states $|\Psi_{0}^{A}\rangle$ are simultaneous eigenstates of these operators.
For a suitable generalization of the model to a surface of genus $g$, this would lead to $4^{g}$ topologically distinct ground
states, one for each sector.

In the following section, we turn to the physical consequences of this
exactly solvable structure, including the locality of
correlations in the
ground state, and the nature of excitations. 

\section{Ultra-local correlations and localized visons}\label{section_corrgap}
Previous work~\cite{wildeboer2020checkerboard} has analytically demonstrated, using a Kasteleyn approach applied to the corresponding classical dimer ensemble obeying the same local constraints, that dimer--dimer correlations on the checkerboard lattice vanish identically whenever the associated unit cells (see Fig.~\ref{figure_666}) do not share a boundary. Since the RK ground state~\eqref{wf} assigns equal weights to all valid dimer coverings, its equal-time correlations coincide with those of this classical ensemble and are therefore ultra-local. Although the checkerboard lattice is non-bipartite, the specific local constraints imposed on crossed plaquettes render the classical
problem amenable to a Kasteleyn treatment. The resulting ultra-short-ranged correlations are closely analogous to those found in the kagome quantum dimer model, where Misguich et. al \cite{misguich} showed directly from the arrow representation that connected local correlators vanish outside a single elementary cell; the same argument applies here as well, as we will show below. Such behavior is a characteristic of topologically ordered systems with local constraints~\cite{Baskaran,Chen}.

To see the ultra-short-rangedness using the arrow representation, one simply notes that arrow variables on distinct sites are statistically independent in the RK ground state unless they participate jointly in a local constraint, i.e., unless they lie on a common crossed plaquette.  
Consequently, any two dimers (or arrows) separated by more than the diameter of a single crossed plaquette share no constraints in common, and their connected correlation function vanishes identically. 

For completeness, we briefly describe the construction of the
elementary excitations of the model.
In the presence of  periodic boundary conditions, the commuting operators $\sigma^{x}(\square)$ are not fully independent. Because every arrow belongs to exactly two uncrossed plaquettes, the product over all such plaquettes, $\prod_{\square}\sigma^{x}(\square)$, flips every arrow twice and therefore acts as the identity on all physical arrow (or dimer) configurations.  Hence, for PBCs the physical Hilbert space is restricted by the global constraint
\begin{equation}
\prod_{\square} \sigma^{x}(\square) = 1\,.
\label{eq:global_constraint_PBC}
\end{equation}
This enforces the constraint that ``visons''~\cite{senthil2000} with $\lambda_{\square}=-1$ can only appear in pairs. No such constraint exists for open boundary conditions, where arrows on the boundary belong to only one plaquette and the product above no longer collapses to the identity.

A vison excitation of an uncrossed plaquette $\alpha$ corresponds to flipping its eigenvalue to $\lambda_\alpha=-1$, which costs an energy $2\gamma$. Because of the global constraint~\eqref{eq:global_constraint_PBC}, a single vison cannot be created on a closed surface; the lowest-energy excitation consists of a \emph{pair} of visons at plaquettes $\alpha$ and $\beta$, with total energy $E_{\Delta}=4\gamma$. To describe these excitations it is convenient to use the dimer representation of Sec.~\ref{section_QDM_model}, where each operator $\sigma^{x}(\square)$ acts as a sum over resonance moves $T_C$ supported on the corresponding superplaquette via Eq.~\eqref{duality}. 

Let $\Omega(\alpha,\beta)$ be a string operator measuring the $\mathbb{Z}_{2}$ parity of dimers intersecting a path connecting $\alpha$ and $\beta$ (avoiding vertices). This can be interpreted as a 't Hooft (dual Wilson) line, especially in the context of Sec.~\ref{section_Ising}. The string operator $\Omega(\alpha,\beta)$ commutes with all $\sigma^{x}(\square)$ except at the
endpoints:
\begin{equation}
\sigma^{x}(\square)\,\Omega(\alpha,\beta)
= \begin{cases}
-\Omega(\alpha,\beta)\,\sigma^{x}(\square), & 
\square=\alpha \text{ or } \beta,\\[4pt]
\;\;\Omega(\alpha,\beta)\,\sigma^{x}(\square), & \text{otherwise}\,.
\end{cases}
\end{equation}
This is so because the string defining $\Omega(\alpha,\beta)$ 
intersects any resonance loop associated with any superplaquette other than those associated to $\alpha$, $\beta$ an even number of times, and the dimer presence/absence at an intersection is changed during any resonance moves. The above equation then follows because the local resonance loops around $\alpha$ and $\beta$, unlike all others, will be intersected by the string an odd number of times.
Hence, acting with $\Omega(\alpha,\beta)$ on a ground state flips $\lambda_{\alpha}$ and $\lambda_{\beta}$ to $-1$, creating a vison pair: 
\begin{equation}
|\Psi^{\rm vis}_{\alpha,\beta}\rangle =
\Omega(\alpha,\beta)\,|\Psi_{0}^{A}\rangle\, ,
\end{equation}
which is an exact eigenstate of energy $4\gamma$ relative to the ground state. 
In particular, the excitation energy is independent of the length or shape of the string.

More generally, products of string operators $\Omega(\alpha,\beta)$ may be used to create arbitrary collections of vison pairs. Open strings terminate on visons and therefore generate excited states, whereas closed strings carry no endpoints. For a contractible closed path, the corresponding operator reduces to the parity of the number of enclosed sites and therefore acts trivially within the constrained Hilbert space. By contrast, closed strings winding nontrivially around the torus define topological operators that distinguish the different ground-state sectors. These are precisely the winding-number operators $\Omega_{x}$ and $\Omega_{y}$ introduced in Sec.~\ref{section_QDM_model}.

\section{Ising ($\mathbb{Z}_2$) gauge theory}\label{section_Ising} 

The generalized arrow representation introduced in Sec.~\ref{section_arrow_ham} naturally suggests a formulation in terms of a $\mathbb{Z}_{2}$ lattice gauge theory. Indeed, such gauge-theory descriptions arise naturally in quantum dimer models defined on medial lattices. In the present case, the arrow variables provide a bridge between dimer coverings and Ising gauge degrees of freedom. In the checkerboard construction presented above, the relevant gauge theory lattice is the square lattice formed by the centers of the crossed plaquettes. For brevity, we will refer to the latter just as the ``gauge lattice''. The medial lattice of this gauge lattice is again a square lattice, more precisely, it is the square lattice underlying the original checkerboard lattice (Fig.~\ref{figure5}). Although this medial lattice, strictly speaking, does not include the crossed links of the checkerboard lattice, the medial lattice naturally has two types of plaquettes: those centered on the vertices on the gauge lattice, and those centered on its plaquettes. It is natural for these two types of plaquettes, though both are square, to play different roles in the resulting QDM. 
In our framework, this is manifest by the introduction of the crossed links meeting in the vertices of the gauge lattice.

As we will see in greater detail in Sec.~\ref{section_ruby}, this construction generalizes to gauge theory/QDM pairs, where the gauge theory lives on any tetravalent lattice. This, in turn, may be seen as a generalization of the framework established by Misguich, Serban, and Pasquier~\cite{misguich} for trivalent gauge lattices, which includes the honeycomb/kagome pair. 
Note that in contrast with ours, their construction does not involve any non-planar lattice graphs.

Moreover, in the present, tetravalent scheme, the standard $\mathbb{Z}_{2}$ lattice gauge theory on the \emph{square lattice} becomes an admissible parent theory for a QDM. It is interesting that the rather intricate quantum dimer model defined above reduces, upon dualizing, to a $\mathbb{Z}_{2}$ gauge theory in the most familiar geometric setting—identical to the setting of Wegner’s original pure $\mathbb{Z}_{2}$ lattice gauge  theory~\cite{Wegner}, and that used by Fradkin and Shenker~\cite{FradkinShenker}, who further studied coupling to dynamical matter fields. This not only makes the connection between the dimer model and conventional lattice  gauge theory particularly transparent, but also renders the checkerboard QDM explicitly equivalent to the toric code~\cite{kitaev2003fault}.

\subsection{Gauge variables and Gauss' law}\label{section_gauge}

We begin by observing that each arrow in the original checkerboard model sits at the midpoint of a link connecting two neighboring sites of the gauge lattice. We therefore associate with each gauge lattice link $\ell$ a pair of Pauli operators $(\tau_\ell^x,\tau_\ell^z)$
acting on a two-dimensional local Hilbert space.

In the original arrow description, we denote by $\sigma_{i}^{z}$ the operator that measures the local arrow orientation at site $i$ 
relative to a reference arrow-state/dimer-state $|D_{\rm ref}\rangle$. 
We now identify the $\mathbb{Z}_{2}$ gauge variables on the dual link $\ell$ whose midpoint coincides with $i$ via
\begin{equation}
\tau_\ell^z \;\equiv\; \sigma_i^x,
\qquad
\tau_\ell^x \;\equiv\; \sigma_i^z\,.
\label{eq:tau_dictionary}
\end{equation}
Note that the choice of reference state amounts to a gauge choice and does not affect any physical result.

The local arrow constraint on each crossed plaquette—the requirement that an even number of arrows point inward—can be expressed in terms of the $\sigma_i^z$ operators and hence in terms of the $\tau_\ell^x$.  
A crossed plaquette corresponds to a gauge lattice site $s$, and its four corner arrows correspond to the four incident gauge lattice links $\ell\ni s$. The even-parity condition is then equivalent to the $\mathbb{Z}_2$ Gauss law
\begin{equation}
G_s \;=\; \prod_{\ell \ni s} \tau_\ell^x,
\qquad
G_s\,|\Psi\rangle = |\Psi\rangle\, ,
\label{eq:GaussLaw}
\end{equation}
for every gauge lattice site $s$. Thus the physical Hilbert space of arrow (or dimer) configurations is precisely the gauge-invariant subspace of a  $\mathbb{Z}_{2}$ lattice gauge theory on a square lattice.

\subsection{Magnetic flux operators}

Given the identification~\eqref{eq:tau_dictionary}, the plaquette operator $\sigma^{x}(\square)$ of Sec.~\ref{section_arrow_ham}—which flips the four  arrows at the corners of an uncrossed plaquette $\square$—becomes the product  of the corresponding $\tau^z$ operators around the boundary of the associated  gauge lattice plaquette $\square^\star$:
\begin{equation}
\sigma^{x}(\square)
\;=\;
\prod_{\ell\in \partial \square^\star} \tau_\ell^z
\;\equiv\;
B_{\square^\star}\,.
\label{eq:flux_operator}
\end{equation}
Thus $\sigma^{x}(\square)$ is exactly the $\mathbb{Z}_{2}$ magnetic-flux operator on the dual lattice.

The exactly solvable Hamiltonian~\eqref{hamil},
\[
{\cal H} = -\gamma \sum_{\square} \sigma^{x}(\square)
= -\gamma \sum_{\square^\star} B_{\square^\star},
\]
is therefore a pure-flux Hamiltonian at the deconfined $\mathbb{Z}_{2}$ 
Rokhsar--Kivelson point. The ground states obtained in Sec.~\ref{section_arrow_ham} are precisely the  flux-free states satisfying $B_{\square^\star}=+1$ for every uncrossed plaquette.

\subsection{Wilson lines, dual strings, visons, and degeneracy}
\label{section_wilson_vison}

In the $\mathbb{Z}_2$ gauge theory language introduced above, it is useful to distinguish between Wilson lines, built from the gauge fields $\tau^z_\ell$, and their dual counterparts, which create magnetic-flux excitations (visons).

A Wilson line is defined as the product of $\tau^z_\ell$ operators along a path $\Gamma$ consisting of links of the gauge lattice:
\begin{equation}
W(\Gamma) \;=\; \prod_{\ell \in \Gamma} \tau^z_\ell\, .
\end{equation}
For a closed path $\Gamma$, this operator is gauge invariant.
Indeed, under a local gauge transformation generated by $G_s$, exactly two links of any closed path touch a given site $s$, so that the two factors of $\tau^z_\ell$ picked up at that site lead to cancellation.
The eigenvalues of noncontractible Wilson loops label the distinct topological sectors of the theory on a surface of nonzero genus, in agreement with the ground state degeneracy discussed earlier.

In $\mathbb{Z}_{2}$ gauge theory, vison excitations naturally correspond to plaquettes of the gauge lattice carrying a nontrivial $\mathbb{Z}_{2}$ magnetic flux, $B_{\square^\star}=-1$.
Such flux excitations are created in pairs by string operators dual to Wilson lines.  We introduce a dual string (’t~Hooft) operator $\tilde W(\alpha,\beta)$ connecting two plaquettes $\alpha$ and $\beta$ of the gauge lattice (equivalently, two uncrossed plaquettes of the checkerboard lattice), defined by its action on the gauge fields as
\begin{equation}
\tilde W(\alpha,\beta) \;=\;
\prod_{\ell^\star \in \Gamma^\star_{\alpha\beta}} \tau^x_{\ell^\star}\, ,
\label{eq:tHooft_def}
\end{equation}
where $\Gamma^\star_{\alpha\beta}$ is a path on the lattice dual to the gauge lattice connecting $\alpha$ and $\beta$.

By construction, $\tilde W(\alpha,\beta)$ commutes with all magnetic-flux operators $B_{\square^\star}$ except for those at its endpoints, viz. 
\begin{equation}
\sigma^{x}(\square)\,\tilde W(\alpha,\beta)
=
\begin{cases}
-\,\tilde W(\alpha,\beta)\,\sigma^{x}(\square),
& \square=\alpha \text{ or } \beta, \\[4pt]
\;\;\tilde W(\alpha,\beta)\,\sigma^{x}(\square),
& \text{otherwise}.
\end{cases}
\label{eq:tHooft_comm}
\end{equation}
\begin{figure}[b]
\includegraphics[width=1.00\columnwidth]{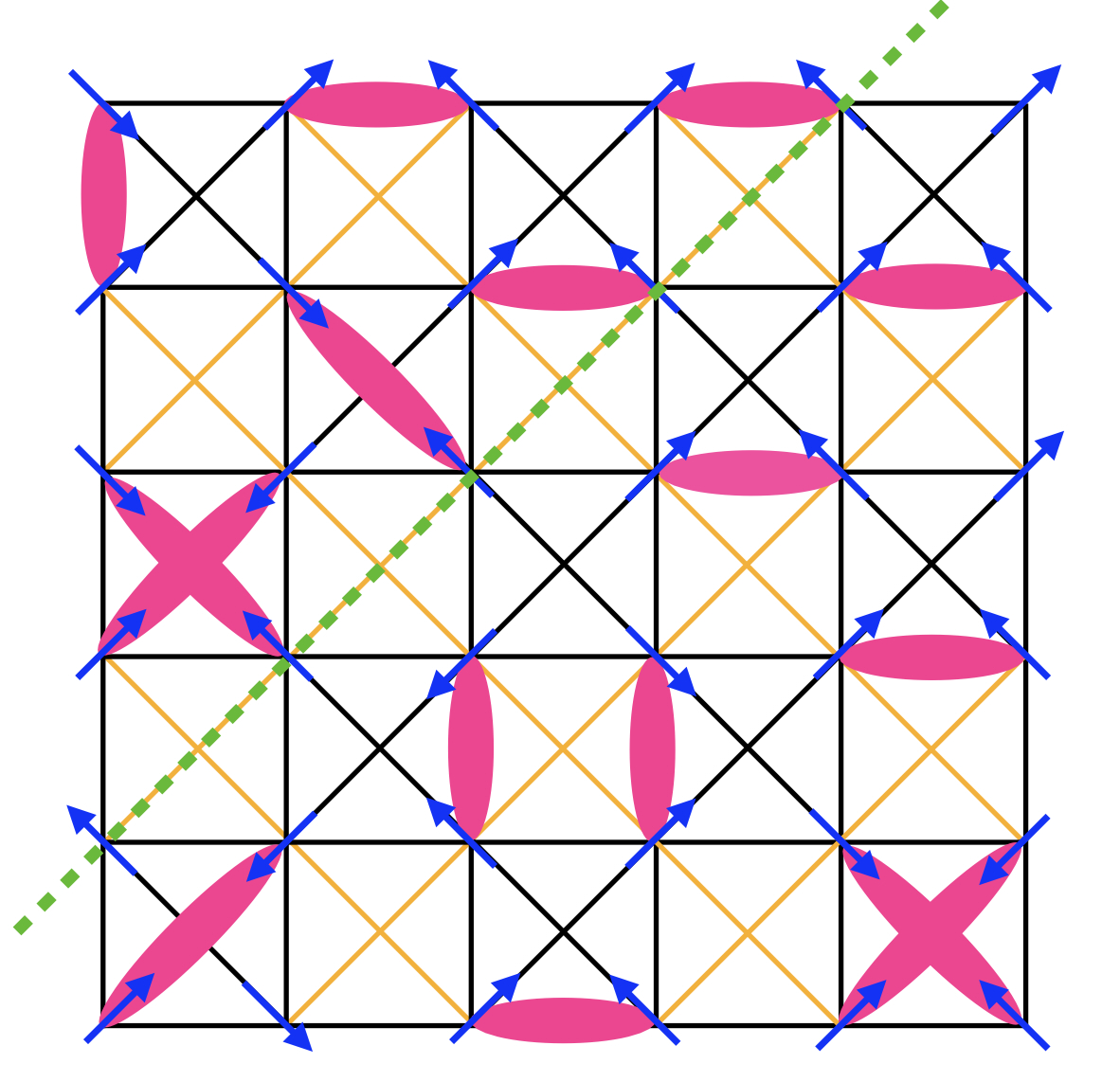}
\caption{
         {\it Reference dimer configuration and associated gauge-theory construction.}
         Shown is an allowed checkerboard dimer configuration together with its associated arrow variables (blue arrows). The black diagonal links of the crossed plaquettes delineate the square lattice on which the $\mathbb{Z}_{2}$ gauge theory lives. The checkerboard lattice is the medial lattice of this gauge lattice, while the yellow lattice denotes the dual of the gauge lattice. The dashed diagonal line indicates a cut running along links of the dual of the gauge lattice, as used later in the discussion of bipartite entanglement. 
}
\label{figure5}
\end{figure}

As a result, $\tilde W(\alpha,\beta)$ has the same algebraic properties as the operator $\Omega(\alpha,\beta)$ introduced earlier in the dimer representation. Indeed, when restricted to the physical (gauge-invariant) Hilbert space, the two operators act identically and may be identified,
\begin{equation}
\tilde W(\alpha,\beta) \;\equiv\; \Omega(\alpha,\beta)\,.
\end{equation}
This identification holds despite the fact that the two operators are defined using slightly different microscopic paths: $\Omega(\alpha,\beta)$ is defined along paths that avoid original lattice sites, whereas $\tilde W(\alpha,\beta)$ is defined along paths that pass through them.  This difference reflects the use of different microscopic degrees of freedom (dimers versus arrows) in their
respective definitions.  In both cases, however, the paths begin and end on uncrossed plaquettes (plaquettes of the gauge lattice), and it is this endpoint structure—rather than the detailed choice of path—that carries physical meaning.

It is instructive to also identify the Wilson loop operators $W(\Gamma)$ in the dimer picture.  Since $W(\Gamma)$ is defined as a product of $\tau^z$ operators—and hence of the arrow-flipping operators $\sigma^x_i$—its action is to flip arrows along the path $\Gamma$.  In the dimer representation, this corresponds to performing a resonance move along a loop $\Gamma'$ that passes through the crossed plaquettes touched by $\Gamma$.  While $\Gamma'$ runs along links of the
original checkerboard lattice and its precise geometry depends on the
instantaneous dimer configuration, the path $\Gamma$ uniquely specifies the set of participating crossed plaquettes. In particular, the magnetic flux terms $B_{\square^\star}=\sigma^x(\square)$ are simply
the shortest possible Wilson loop operators.  More generally, Wilson loops may execute extended resonance processes, including those associated with noncontractible cycles on surfaces of nonzero genus.

In Secs.~\ref{section_QDM_model} and ~\ref{section_corrgap}, we discussed topological degeneracy in terms of the dual (t’Hooft) operators $\Omega$ or $\tilde W$ along loops that are generators of the fundamental group.  An equivalent characterization may be given in terms of Wilson loops. These Wilson loops and their duals form mutually anticommuting pairs, and the ground state degeneracy is given by the dimension of the smallest representation of the resulting algebra.

In closing this section, we note that the model discussed here corresponds to a pure-flux $\mathbb{Z}_{2}$ gauge theory. A natural extension is obtained by adding a transverse-field term for the gauge variables,
$-h\sum_{\ell}\tau^x_{\ell}$, which introduces dynamics for the
$\mathbb{Z}_2$ electric field. Such a term may be viewed as creating or
annihilating pairs of visons on neighboring plaquettes and leads to a
standard interacting variant of  $\mathbb{Z}_2$ lattice gauge theories~\cite{Wegner,FradkinShenker,Kogut}.  While this deformation destroys exact solvability, it does not alter the structure of the deconfined phase realized at small $h$, and we do not pursue it further here.

\section{Entanglement entropy and minimum entropy states}\label{section_ee}

Entanglement entropy affords fundamental insights for probing the topological nature of quantum systems. While the preceding section already establishes the quantum dimer model~\eqref{eq:dimerHamiltonian_final} as belonging to the $\mathbb{Z}_{2}$ topological phase, one advantage of exactly solvable models is that they provide a testbed for the exact calculation of characteristic signatures of the phase.
Exact entanglement properties of the pure-flux $\mathbb{Z}_{2}$ gauge theory on the square lattice, or equivalently, the toric
code, are well established~\cite{zhang2012quasiparticle}. In spite of the one-to-one correspondence, translation of these properties to the quantum dimer models requires some care. This is so since the relationship between the dimers and arrows of our model is not entirely local. Problems stemming from this non-locality can be avoided by considering bipartitions along cuts of the kind shown in Fig.~\ref{figure5}. That is, we cut along diagonals of the uncrossed plaquettes, or, equivalently, along links of the dual of the 
gauge lattice. The advantage of such cuts is that the arrows along the cuts are not independent degrees of freedom, but can be determined from the arrows on either side, {\it as well as} from the dimers on either side. This preserves, in particular, a one-to-one correspondence between dimer and arrow degrees of freedom {\it on either side} of the cut. Thus, for such cuts, questions relating to bipartite entanglement are identical for the dimer model and the $\mathbb{Z}_{2}$ gauge theory.
In particular, the standard results for topological
entanglement entropy (TEE), minimum entropy states (MES), and modular
matrices derived for the toric code apply directly
to the present model~\cite{LW,KP,zhang2012quasiparticle}.  We briefly
summarize these results in the language of the checkerboard dimer model.

\begin{figure}
\includegraphics[width=1.00\columnwidth]{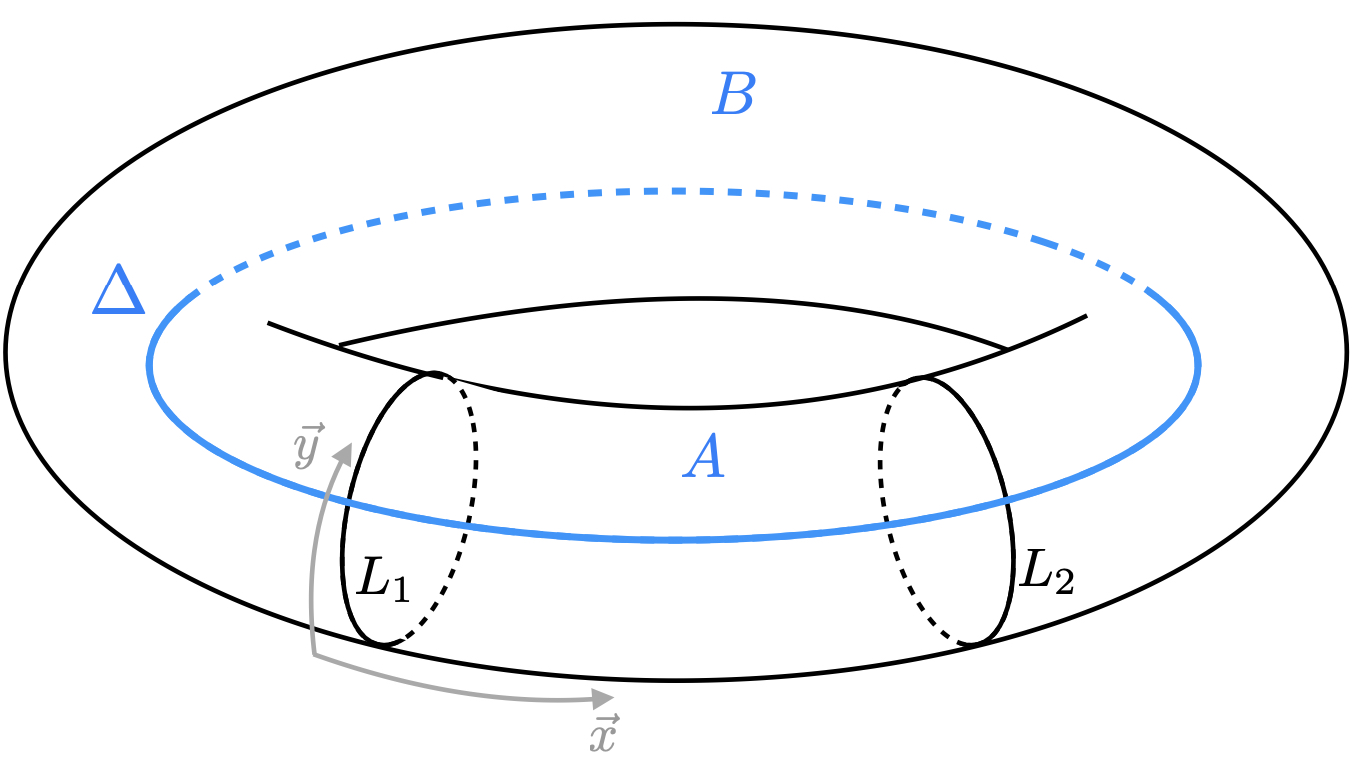}
\caption{
        {\it Cylindrical bipartition of the torus used for entanglement calculations.}
        A torus bipartitioned into two cylindrical regions $A$ and $B$ by two entanglement boundaries wrapping around the $\hat y$-direction. The auxiliary noncontractible cut $\Delta$, running along the $\hat x$-cycle of the torus, is used in the Schmidt decomposition discussed in Sec.~\ref{section_ee} and in the construction of the minimum-entropy states.         
}
\label{figure6}
\end{figure}

We will work on a checkerboard torus obtained by imposing standard periodic boundary conditions on the gauge lattice. We bipartition the torus into two cylindrical regions $A$ and $B$
[see Fig.~\ref{figure6}], along cuts as described above, and consider the $n$-th Renyi entanglement entropy
\begin{eqnarray}\label{renyi}
S_n=\frac{1}{1-n}\ln \mathrm{Tr}(\rho_A^n)\, ,
\end{eqnarray}
where $\rho_{A}=\mathrm{Tr}_{B} |\Psi\rangle\langle\Psi|$ is the reduced
density matrix of region $A$. Ground states of local Hamiltonians obey
an area law of the form~\cite{Eisert}
\begin{eqnarray}\label{S2}
S_n(\rho_A)=\alpha_n L_A-\gamma+\cdots\, ,
\end{eqnarray}
where $L_{A}$ is the length of the entanglement boundary and
$\gamma=\ln(\mathcal D)$ is the TEE determined by the total quantum
dimension $\mathcal D = \sqrt{\sum_{i} d_{i}^{2}}$ defined through the quantum dimensions of the individual quasiparticles $d_{i}$~\cite{LW,KP,dong2008topological}. Conventionally ordered phases have $\mathcal D = 1$, while topologically ordered phases have $\mathcal D > 1$. For the $\mathbb{Z}_2$ topological phase realized here, one has $\mathcal{D}=2$ and therefore $\gamma=\ln 2$.

For bipartitions with noncontractible boundaries, such as the
two-cylinder cut of Fig.~\ref{figure6}, the subleading constant becomes
state dependent. Expressing a generic ground state $|\Psi_\alpha\rangle$ in a MES basis
$\{|\Xi_j\rangle\}$,
\begin{equation}
|\Psi_\alpha\rangle=\sum_j c_j |\Xi_j\rangle ,
\end{equation}
the constant contribution to the Renyi entropy is~\cite{zhang2012quasiparticle}
\begin{eqnarray}\label{gamma_torus2}
\gamma_n'(\{p_j\}) = 2\gamma +
\frac{1}{n-1}
\ln
\left(
\sum_j p_j^n d_j^{2(1-n)}
\right),
\end{eqnarray}
where $p_j=|c_j|^2$ and $d_j$ denotes the quantum dimension of the
$j$-th anyon type. 

The checkerboard model possesses four topological sectors labeled by the
parities of dimer crossings along the two noncontractible directions of
the torus. Following the notation of
Sec.~\ref{section_QDM_model}, the sector ground states are
\begin{eqnarray}\label{state_ab}
| \Psi_{ab} \rangle =
\frac{1}{\sqrt{{\mathcal N}_{ab}}}
\sum_{D\in(ab)} |D\rangle 
\end{eqnarray}
with $(ab)\in\{00,01,10,11\}$. 
Here, $\mathcal N_{ab}$ denotes the number of dimer coverings in the
corresponding sector,
in terms of which we write the
general ground state as the superposition
\begin{eqnarray}\label{sector_all}
| \Psi \rangle = \sum_{(ab)} c_{ab} |\Psi_{ab}\rangle \,.
\end{eqnarray}

Proceeding as in Ref.~\cite{zhang2012quasiparticle}, one performs a
Schmidt decomposition with respect to the cylindrical bipartition.
This yields 
the following basis of
MES states 
\begin{align}
\begin{split}\label{MES_Z2_supp}
| \Xi_{1}^{\vec{x}}\rangle
&=
\frac{1}{\sqrt{2}}
\left(
| \Psi_{00} \rangle + | \Psi_{01} \rangle
\right)\\
| \Xi_{2}^{\vec{x}}\rangle
&=
\frac{1}{\sqrt{2}}
\left(
| \Psi_{00} \rangle - | \Psi_{01} \rangle
\right)\\
| \Xi_{3}^{\vec{x}}\rangle
&=
\frac{1}{\sqrt{2}}
\left(
| \Psi_{11} \rangle + | \Psi_{10} \rangle
\right)\\
| \Xi_{4}^{\vec{x}}\rangle
&=
\frac{1}{\sqrt{2}}
\left(
| \Psi_{11} \rangle - | \Psi_{10} \rangle
\right)\,.
\end{split}
\end{align}
With the above, we may write the probabilities entering
Eq.~\eqref{gamma_torus2} as
\begin{align}
\begin{split}\label{4pj}
p_{1} &= \frac{|c_{00}+c_{01}|^{2}}{2},
\qquad
p_{2} = \frac{|c_{00}-c_{01}|^{2}}{2},
\\
p_{3} &= \frac{|c_{10}+c_{11}|^{2}}{2},
\qquad
p_{4} = \frac{|c_{10}-c_{11}|^{2}}{2}.
\end{split}
\end{align}
For the individual topological-sector states $|\Psi_{ab}\rangle$, one obtains $\gamma'=\ln 2$, while for the MES states $|\Xi_i\rangle$ one finds $\gamma'=2\ln 2$, consistent with the MES entanglement signatures previously obtained for the toric code~\cite{zhang2012quasiparticle} and the kagome lattice quantum dimer model~\cite{wildeboer2017entanglement}.

The MES states furthermore determine~\cite{zhang2012quasiparticle} the modular $\mathcal S$-matrix through
\begin{eqnarray}\label{S_matrix}
\mathcal{S}_{\alpha\beta}
=
\frac{1}{\mathcal D}
\langle
\Xi^{\vec x}_\alpha
|
\Xi^{\vec y}_\beta
\rangle\,.
\end{eqnarray}
Using $d_{i}=1$ for all individual quasiparticles $i=1,\ldots,4$, one obtains $\mathcal{D}=2$ and the $\mathcal{S}$-matrix 
\begin{align}\label{Smatrix}
\mathcal{S}
=
\frac{1}{2}
\begin{pmatrix}
 1 & \,\,\,\,\,1 & \,\,\,\,\,1 & \,\,\,\,\,1 \\
 1 & \,\,\,\,\,1 &          -1 &          -1 \\
 1 &          -1 & \,\,\,\,\,1 &          -1 \\
 1 &          -1 &          -1 & \,\,\,\,\,1
\end{pmatrix}
\end{align}
corresponding to the modular data of the $\mathbb{Z}_2$ topological phase. 
\begin{figure*}[t]
\includegraphics[width=1.00\textwidth]{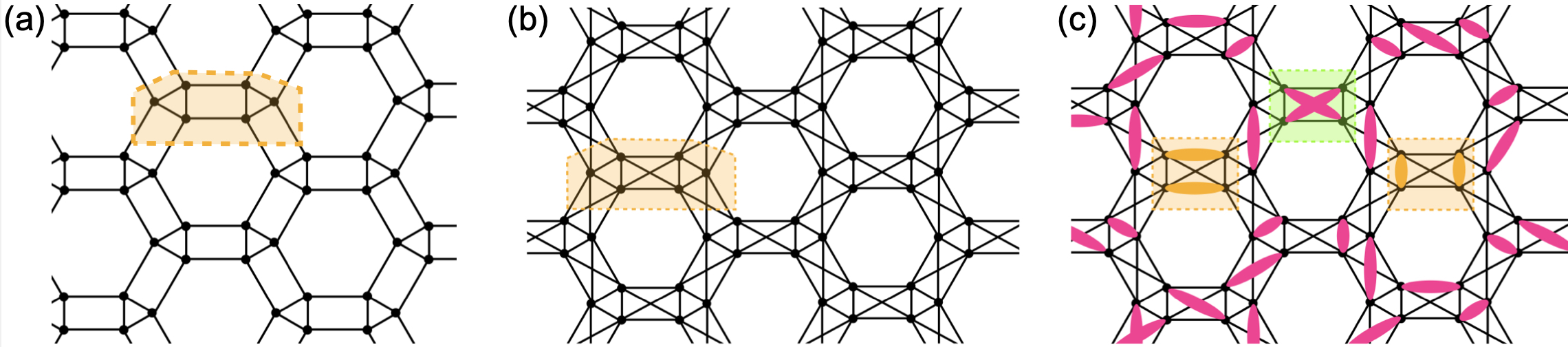}
\caption{
        {\it Ruby-lattice generalization and local crossed-plaquette constraint.}
        (a) Regular ruby lattice consisting of two plaquette types, triangles and rectangles. The highlighted rectangle indicates a representative four-site plaquette of the type that plays the role of a crossed plaquette in the checkerboard construction, but here without diagonal bonds. 
        (b) Crossed ruby lattice obtained by inserting two diagonals into each rectangular plaquette, thereby producing crossed plaquettes. 
        The diagonal links of the rectangular plaquettes intersect at the sites of an underlying kagome lattice, which plays the role of the parent IGT lattice in the present construction. Equivalently, the ruby lattice is the medial lattice of the kagome lattice. The highlighted rectangle indicates the corresponding local motif. 
        (c) Local dimer constraint on the crossed ruby lattice. Whenever two dimers occupy a crossed plaquette, they must form a crossed pair on the two diagonal links (green, allowed), while parallel two-dimer occupancies on the same plaquette are forbidden (orange).
}\label{fig:ruby7}
\end{figure*}

\section{Ruby lattice version of the model}\label{section_ruby}

We now show that our checkerboard construction admits generalization to
other lattices, notably the ruby lattice. The essential structural ingredient underlying the solvability mechanism is the presence of corner-sharing crossed {\it rectangular} plaquettes, sharing edges with
uncrossed plaquettes. More precisely, for the map to a dual Ising gauge theory to exist, one needs the dimer lattice to be the medial lattice of the gauge theory lattice. This was already realized as the universal construction principle of the kagome lattice QDM, where the gauge theory lattice must be trivalent~\cite{misguich}. We will show here that our construction generalizes this principle to tetravalent lattices. That is, the most general lattice for which the construction of our QDM works is the medial lattice of a tetravalent lattice. This is the reason why the square lattice/checkerboard lattice fits nicely into our paradigm. Moreover, it makes the ruby lattice an obvious next candidate. This is furthermore motivated by recent work~\cite{Verresen2021, Verresen2022}, which has identified ruby-lattice models supporting $\mathbb{Z}_{2}$ topological order in related constrained settings. Our construction naturally results in a complementary exactly solvable instance where the same phase emerges.

In addition to providing the topological ingredient for a dual IGT to be formulated, we will show below that the medial lattice construction, in particular, guarantees the ergodicity of the dimer moves (within topological sectors).
We now focus on the ruby lattice, but keep the discussion general enough such that generalization to any medial lattice of a tetravalent lattice will be obvious. Interestingly, our generalization of the original trivalent medial construction (the original honeycomb IGT/kagome QDM paradigm) always leads to QDMs with crossed diagonals, hence, non-planar graphs and dimer configurations with intersecting loops in their transition graphs, counter to the established template for solvable QDMs. 

The ruby lattice is the medial lattice of the tetravalent kagome lattice (Fig.~\ref{fig:ruby7} and caption). Plaquettes of the ruby lattice come in two types: Plaquettes associated with the original kagome lattice sites -- these must be rectangular and corner-sharing owing to the tetravalent nature of the kagome lattice. We will adorn these rectangular plaquettes with diagonal links, leading to a ``crossed'' ruby lattice.
The other types of plaquettes (triangles and hexagons) are associated with the plaquettes of the original kagome lattice (also triangles and hexagons), and remain uncrossed. These types of plaquettes generalize the crossed and uncrossed squares of our checkerboard construction.

A useful geometric perspective is that the diagonal dressing introduced in the present construction directly delineates the underlying parent lattice that hosts the IGT, in the present case the kagome lattice.
The dimers of the QDM then live both on the links of the medial lattice and on the added diagonal links, whose connectivity mirrors that of the parent gauge lattice itself (although the corresponding links have
different endpoints in the two constructions). In this sense, the crossed-medial QDMs studied here may be viewed as hybrid structures combining the medial lattice with the connectivity skeleton of the underlying IGT lattice. 

In the following Sec.~\ref{rubyA}, we will discuss the construction of an exactly solvable commuting-projector Hamiltonian on the ruby lattice, emphasizing those aspects that require generalization from the simplest square/checkerboard case. Subsequently, in Sec.~\ref{rubyB}, we show that the standard Kasteleyn/Bloch formulation can also be applied to the crossed ruby lattice and contrast its constrained structure with that of the regular ruby lattice dimer problem.

Figures~\ref{fig:ruby7} and ~\ref{fig:ruby8abc} collect the relevant
lattice geometries, local constraints, superplaquette structures, and
unit-cell conventions, while Appendix~\ref{app:ruby_pfaffian} contains the detailed Kasteleyn derivations and the full cell-indexing conventions used in the Bloch construction. 

\subsection{Regular versus crossed ruby lattice: model, constraints,
and mapping}\label{rubyA}

As motivated above, for the crossed ruby lattice, the role of the crossed plaquettes is played by the four-site rectangular plaquettes carrying two diagonals, while the role of the uncrossed plaquettes is played by the elementary triangles and hexagons [Fig.~\ref{fig:ruby7}]. The local constraint is the direct analogue of the checkerboard case: whenever two dimers occupy a crossed plaquette, they must form a crossed pair, whereas parallel two-dimer occupancies are forbidden.  The allowed and forbidden local configurations are illustrated in Fig.~\ref{fig:ruby7}(c). In the dual IGT picture on the kagome lattice,
the arrow constraint associated with rectangular ruby-plaquettes becomes Gauss' law, whereas the dimer moves around uncrossed plaquettes measure magnetic fluxes.

\begin{figure*}[t]
\includegraphics[width=1.00\textwidth]{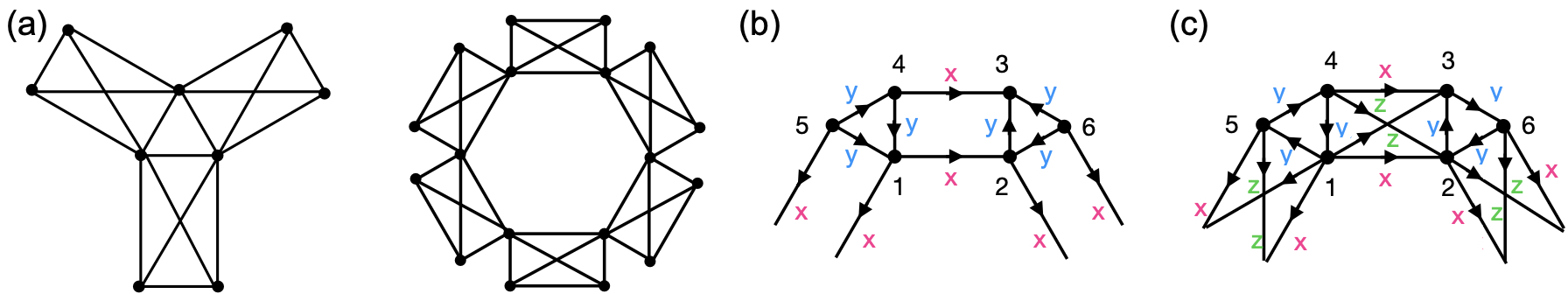}
\caption{ 
        {\it Ruby lattice superplaquette and unit-cell conventions for the Pfaffian/Bloch construction.}
        (a) Ruby-lattice superplaquettes consisting of a central triangle together with the three adjacent crossed plaquettes and a central hexagon together with the six adjacent crossed plaquettes. This is the local object on which the ruby analogue of the checkerboard resonance move acts. 
        (b) Six-site primitive unit cell and a Kasteleyn bond orientation for the regular ruby lattice. Bond weights are labeled by $x$ on the long rectangular bonds (magenta bonds) and by $y$ on the triangle bonds (blue labels).  
        (c) Six-site  primitive unit cell and pre-Kasteleyn bond orientation for the crossed ruby lattice. In addition to the $x$- and $y$-bonds, the diagonal bonds inside the crossed plaquettes carry weight $z$ (green labels). These unit-cell conventions are used throughout Sec.~\ref{section_ruby} and Appendix~\ref{app:ruby_pfaffian}. 
}\label{fig:ruby8abc}
\end{figure*}

The QDM is again defined through the correspondence between arrow and
dimer configurations. In addition, the construction induces a local
mapping between arrow configurations and resonance loops on
superplaquettes consisting of an uncrossed plaquette together with the
surrounding crossed rectangles [Fig.~\ref{fig:ruby8abc}(a)]. More specifically, one may again introduce a binary arrow variable on every lattice site, constrained so that the number of inward-pointing arrows on each crossed plaquette is even, i.e. $n_{\mathrm{in}}=0,2,4$. The uncrossed plaquettes then play the role of the local update units: flipping all arrows around a triangle or hexagon preserves the parity constraint on every adjacent crossed plaquette.  

For any arrow configuration satisfying the local parity constraints, the corresponding dimer configuration on crossed plaquettes is uniquely determined by the same rules introduced for the checkerboard case.
Moreover, this then also defines a closed loop on any superplaquette, using again the same local map between arrows on surrounding crossed rectangles and loop segments.  Flipping the arrows on the central uncrossed plaquette then exchanges occupied and unoccupied links along this loop. 
In the dimer basis, the operator flipping arrows around uncrossed plaquettes therefore becomes a sum of loop-specific resonance moves
supported on the corresponding superplaquette, and the full dimer Hamiltonian is again a purely kinetic RK-point Hamiltonian built from these local resonances. 

In summary, this again leads to the following exactly solvable
commuting-projector Hamiltonian. For each triangle $\triangle$ or hexagon $\hexagon$, we define
\begin{align}
\begin{split}
\sigma^x(\triangle)&=\prod_{i\in\triangle}\sigma_i^x \\
\sigma^x(\hexagon)&=\prod_{i\in\hexagon}\sigma_i^x
\label{eq:ruby_sigma_triangle}
\end{split}
\end{align}
which flip the arrows around the respective triangles/hexagons. 
The ruby-lattice arrow Hamiltonian is therefore 
\begin{equation}
\mathcal{H}_{\mathrm{ruby}}
= -\gamma\sum_{\triangle}\sigma^x(\triangle) -\gamma'\sum_{\hexagon}\sigma^x(\hexagon) 
\label{eq:ruby_arrow_ham}
\end{equation}
and we introduced positive constants $\gamma$, $\gamma'$. By the commuting nature of the operators $\sigma^x(\triangle)$ and $\sigma^{x}(\hexagon)$, i.e. 
\begin{align}
\begin{split}
&\big[\sigma^x(\triangle), \sigma^x(\triangle^{\prime})\big] 
= \big[\sigma^x(\hexagon), \sigma^x(\hexagon^{\prime})\big]
= 0 \\
&\qquad\qquad\,\,\,\,\big[\sigma^x(\triangle), \sigma^x(\hexagon)\big] 
= 0 
\end{split}
\end{align}
$\forall\, \triangle, \triangle^{\prime} \in \{\triangle\}$, and 
$\forall\, \hexagon, \hexagon^{\prime} \in \{\hexagon\}$, 
$\mathcal{H}_{\mathrm{ruby}}$ is again {\it exactly solvable}.  
Four topological sectors can be defined via Wilson loops or their duals in a manner completely analogous to the one discussed for the checkerboard lattice.
For the model to have exactly one ground state per topological sector, we need the dynamics~\eqref{eq:ruby_arrow_ham} to be ergodic in each sector. This is indeed the case, and can be seen from arguments that are standard for IGTs, or, more directly, from counting arguments we give below. Ground states of~\eqref{eq:ruby_arrow_ham} are thus 
simultaneous $+1$ eigenstates of all operators $\sigma^x(\triangle)$ and $\sigma^x(\hexagon)$, and within each topological sector are given by equal amplitude superpositions of dimer configurations.

The excitation structure likewise parallels the checkerboard case. String operators create pairs of visons at their endpoints, and the resulting phase is again a fully gapped $\mathbb{Z}_{2}$ dimer liquid with deconfined fractional excitations, topological ground-state degeneracy, and the same toric-code modular data discussed in Secs.~\ref{section_corrgap}--\ref{section_ee}. In this sense, the checkerboard and crossed ruby constructions provide two geometrically distinct realizations of the same exactly solvable topological dimer-liquid mechanism.

Similarly to the checkerboard case, we may count the number of dimer coverings per topological sector. 
By comparing this number to the number of independent resonance moves, we can  make the case for the ergodicity of the model within topological sectors. Let $N_{s}$ be the number of ruby sites/arrows. The total number of dimer coverings $N_{D}^{\rm total}$ is then 
\begin{eqnarray}
N_{D}^{\rm total} = 2^{N_{s}-(N_{\square}-1)}
\end{eqnarray}
or $2^{N_{s}-N_{\square}-1}$ per topological sector, where we used that $N_{\square}-1$ is the number of independent arrow constraints, with $N_{\square}$ the number of rectangles of the ruby lattice. Similarly let $N_{\triangle}$ and $N_{\hexagon}$ be the number of triangles/hexagons of the ruby lattice.
The number of independent resonance moves is $N_{\triangle}+N_{\hexagon}-1$ (as the application of all moves flips every arrow twice, giving the identity). 
This gives access to $2^{N_{\triangle}+N_{\hexagon}-1}$ states per {\it dynamical} sector. Ergodicity thus comes down to the identity
\begin{equation}\label{eq:ergodic}
N_{s}-N_{\square} = N_{\triangle}+N_{\hexagon} \,.  
\end{equation}
On the other hand, we can express the above through corresponding quantities in the kagome lattice whose medial lattice is the ruby lattice. We have $N_{s} = E$, where $E$ is the number of kagome edges, $N_{\square} = V$, where $V$ is the number of kagome vertices, and $N_{\triangle}+N_{\hexagon}=F$, where $F$ is the number of kagome faces. Then Eq.~\eqref{eq:ergodic} becomes
\begin{equation}
V-E+F=0
\end{equation}
which is just the statement that the Euler characteristic of a toroidal lattice vanishes. The generality of this argument is apparent for all lattices that are the medial lattices of a tetravalent lattice (and applies similarly to the trivalent case). 

\subsection{Regular versus crossed ruby lattice: Kasteleyn formulation and torus counting}\label{rubyB}

We close by briefly contrasting the crossed ruby construction with the
standard dimer problem on the regular ruby lattice, i.e.\ without
diagonal bonds inside the rectangular plaquettes.  While the regular
ruby lattice remains a conventional planar dimer problem admitting a
standard Kasteleyn formulation, it does not possess the local
commuting-projector structure of the crossed ruby model.
Nevertheless, the same Kasteleyn/Bloch machinery can be developed for
both cases, making it instructive to briefly compare the planar and non-planar settings in the following subsection.  In particular,
while the regular ruby lattice exhibits the dispersive structure
characteristic of a conventional planar dimer model, the crossed ruby
case retains the constrained sector structure associated with the
exactly solvable construction. 

A useful way to compare the regular and crossed ruby models is through their Kasteleyn formulations on the torus. For the crossed ruby lattice, this requires some elaboration. By  definition, a Kasteleyn orientation is defined for (locally) planar lattice graphs, utilizing a well-defined notion of faces: every elementary face must be clockwise odd~\cite{kas}. This definition is not applicable to lattice graphs that are not planar even locally, such as our crossed lattices. 
This problem has been circumvented for the classical version of our checkerboard problem in Ref. 
~\cite{wildeboer2020checkerboard},
using the notion of a ``pre-Kasteleyn'' orientation, originally defined 
in Ref.~\cite{wildeboer2020entanglement}. 
The latter can be applied to non-planar graphs that are nonetheless embedded in a 2D surface, and requires that for every contractible non-self-intersecting loop, the parity of clockwise arrows is opposite to the parity of the enclosed sites. It can be seen that for planar graphs, the two notions agree. It was then shown in Ref. 
~\cite{wildeboer2020checkerboard} 
that for the constrained checkerboard dimer problem, the Pfaffian of the pre-Kasteleyn matrix is related to allowed dimer configurations, owing to a balanced partial cancellation between allowed and disallowed configurations on crossed plaquettes. These arguments carry over to the crossed ruby lattice (and the larger class of models defined in this section) in a one-to-one fashion. 

Figs.~\ref{fig:ruby8abc}(b), (c) define a Kasteleyn and pre-Kasteleyn orientation for the regular and crossed ruby lattice, respectively.
The former can be verified using the standard definition. For the latter, we may use the following criterion: We see that for any dilution of the links obtained via removal of any one link from every crossed rectangle, a standard Kasteleyn orientation is obtained. This renders the given orientation pre-Kasteleyn~\cite{wildeboer2020entanglement}. Furthermore,
certain weights $x,y,z$ may be introduced for the various links, as shown in Fig. ~\ref{fig:ruby8abc}(c).
This results in a class of models, at the wave function/partition function level, that smoothly interpolates between the regular and crossed ruby lattice models, the crossed case corresponding to $x=y=z=1$, and the regular case to $x=y=1$, $z=0$. 
The weights and orientations are encoded in a matrix $A$ in the usual fashion.
Since from here on, the formalism is the same for Kasteleyn and pre-Kasteleyn matrices, we will just refer to $A$ as the ``Kasteleyn matrix'' from now on.

For an $M\times N$ array of unit cells, we denote by $\phi$ and $\theta$ the Bloch angles associated with translations in the $M$- and $N$-directions, respectively. The unit-cell conventions used throughout are shown in Figs.~\ref{fig:ruby8abc}(b), (c), while the full cell-indexing and neighbor-cell conventions entering the Bloch construction are given in Appendix~\ref{app:ruby_pfaffian}. For a given spin structure $(\mu,\nu)\in\{0,1\}^2$, corresponding to periodic or antiperiodic twists along the two torus cycles,
these angles take the standard discrete values
\begin{align}
\begin{split}
\phi&=\phi_{m,\nu}=\frac{2\pi}{M}\Bigl(m+\frac{\nu}{2}\Bigr)\\[3pt]
\theta&=\theta_{n,\mu}=\frac{2\pi}{N}\Bigl(n+\frac{\mu}{2}\Bigr)
\label{eq:ruby_main_sector_angles}
\end{split}
\end{align}
with $m=0,\dots,M-1$ and $n=0,\dots,N-1$. 

For the regular ruby lattice, the Kasteleyn treatment is entirely conventional. Using the six-site unit cell of Fig.~\ref{fig:ruby8abc} (b), one obtains a $6\times6$ Bloch matrix whose determinant is a genuinely dispersive trigonometric polynomial, cf. Eq.~\eqref{eq:A_reg_det_iso}. The sector determinants are then given by Brillouin-zone products over the shifted momentum grids~\eqref{eq:ruby_main_sector_angles}, and the torus partition sum is assembled from the four sectors in the standard way. Thus, the regular ruby lattice behaves as a conventional dispersive classical dimer model, even though its partition function and correlation functions remain efficiently computable through the Kasteleyn/Pfaffian construction. The explicit primitive blocks, Bloch kernel, and sector-product formulas are collected in Appendix~\ref{app:regular_ruby}.

The crossed ruby lattice is qualitatively different. As explained in Sec.~\ref{rubyA}, imposing the crossed-plaquettes constraint yields an exactly solvable commuting-projector quantum dimer model with the same $\mathbb{Z}_{2}$ liquid structure as in the checkerboard case. At the level of the Kasteleyn formulation, the corresponding primitive $6\times6$ Bloch kernel $A_{6}(\theta,\phi)$ is not generically skew-symmetric at fixed $(\theta,\phi)$. Rather, it satisfies the Bloch-skew relation
\begin{equation}
A_6(\theta,\phi)^T=-A_6(-\theta,-\phi)
\label{eq:ruby_main_bloch_skew}
\end{equation}
so that the natural primitive per-mode object is the determinant $\det A_{6}(\theta,\phi)$, rather than a mode-wise Pfaffian.

At the isotropic point
\begin{equation}
t\equiv x=y=z,
\label{eq:ruby_main_isotropic}
\end{equation}
the primitive determinant simplifies dramatically. The calculation in Appendix~\ref{app:spec_6site} gives
\begin{equation}
\det A_6(\theta,\phi)=64\,t^{6} 
\label{eq:ruby_main_special_flat}
\end{equation}
independent of both Bloch angles. Thus, for the crossed ruby lattice model, the primitive Bloch determinant is completely flat at isotropic link weights.

As a brief application, it is instructive to re-derive the counting of
dimer configurations obtained in Sec.~\ref{rubyA}. For the full torus problem, each twist sector $(\mu,\nu)$ is obtained by evaluating Eq.~\eqref{eq:ruby_main_special_flat}
on the shifted momentum grids~\eqref{eq:ruby_main_sector_angles}. Since the corresponding real-space Kasteleyn matrices are antisymmetric, the torus partition sum is assembled from the four Pfaffians in the standard manner, i.e. 
\begin{align}
\begin{split}
&Z_{\mathrm{torus}}(M,N)=\\
&\qquad\quad
\frac12\Big[
-{\rm Pf}(A_{00})
+{\rm Pf}(A_{01})
+{\rm Pf}(A_{10})
+{\rm Pf}(A_{11})
\Big]\,.
\label{eq:ruby_main_special_pf}
\end{split}
\end{align} 
For the ribbon/orientation convention used here, the Pfaffian branches
are chosen such that ${\rm Pf}(A_{00})=-\sqrt{\det A_{00}}$, ${\rm Pf}(A_{01})=+\sqrt{\det A_{01}}$, ${\rm Pf}(A_{10})=+\sqrt{\det A_{10}}$ and ${\rm Pf}(A_{11})=+\sqrt{\det A_{11}}$~\footnote{This is the standard torus sign structure in the Pfaffian-sector decomposition. For a closely related explicit sign table, see Table~I of Ref.~\cite{wildeboer2017entanglement}. In the present convention, the
corresponding branch choice is fixed independently by direct
enumeration on the minimal torus.}, 
so that Eq.~\eqref{eq:ruby_main_special_pf} is equivalently an all-plus sum over the positive square roots of the sector determinants. 

At isotropic link weights, the flat primitive determinant~\eqref{eq:ruby_main_special_flat} then implies the
simple closed form
\begin{equation}
Z_{\mathrm{torus}}(M,N)=2\times(8t^3)^{MN}\,.
\label{eq:ruby_main_special_torus}
\end{equation}
At unit weights $t=1$, this becomes
\begin{equation}
Z_{\mathrm{torus}}(M,N)=2^{3MN+1}\,.
\label{eq:ruby_main_special_torus_unit}
\end{equation}
Since $N_s=6MN$, this agrees with the exact arrow-state counting
discussed in Sec.~\ref{rubyA}. Details, including sector-restricted counting,
are given in Appendix~\ref{app:ruby_pfaffian}. 

The regular and crossed ruby lattices therefore provide two closely
related but physically distinct realizations of the same underlying
Kasteleyn framework. The regular ruby model behaves as a conventional
dispersive planar dimer system with nontrivial correlations, while the
crossed-ruby construction collapses onto an exactly solvable constrained
$\mathbb{Z}_{2}$ dimer liquid with flat Bloch structure and closed-form
torus counting. The interpolation between the two cases through the
bond-weight parameter $z$ furthermore illustrates how the present
crossed-medial construction naturally extends the conventional planar
Kasteleyn setting into the non-planar pre-Kasteleyn regime.

\section{Conclusion}\label{section_conclusions}

In this work, we generalized an influential framework for the construction of fully solvable quantum dimer models with dual Ising gauge theory descriptions, realizing the $\mathbb{Z}_{2}$ topological phase. While the original framework had IGTs living on general trivalent parent lattices, our construction generalizes this to tetravalent lattices. The resulting quantum dimer model, defined on the medial lattice of the IGT, then has the distinctive feature of having crossed plaquettes, leading to dimer configurations whose transition graphs generically contain intersecting loops. The structural element of non-planarity runs counter to established templates for exactly tractable dimer models.

Our construction in particular applies to the arguably simplest IGT
setting, namely the square lattice, which maps to a QDM on the
checkerboard lattice through the generalized arrow representation that
forms the centerpiece of our approach. Through its dual square-lattice
IGT description, the checkerboard model is in fact exactly equivalent
to the original toric code~\cite{kitaev2003fault}. 
We then showed that a QDM on the ruby lattice, dual to an IGT on the kagome lattice, likewise falls naturally within the scope of the framework. The ruby lattice had recently been identified as a fruitful playground for constrained dynamics supporting $\mathbb{Z}_{2}$  topological order~\cite{Verresen2021, Verresen2022}, and our construction provides an exactly solvable realization within this broader setting.

More generally, our construction gives rise to a broad new class of
exactly solvable crossed-medial quantum dimer models associated with
tetravalent parent lattices. While the original trivalent
medial-lattice construction already generates infinitely many
kagome-type quantum dimer models, the tetravalent setting admits a
natural iterative structure. This follows from the elementary fact that
the medial lattice of any two-dimensional lattice is itself tetravalent. Starting from a tetravalent IGT lattice, one obtains a
crossed-medial QDM whose underlying lattice graph, after removing the
crossed decorations, can again serve as a new admissible IGT geometry.
For example, the crossed ruby model constructed from the kagome lattice
naturally leads back to the regular ruby lattice as the next IGT host;
its medial lattice in turn supports another crossed-medial QDM, and so
on. Iterating this procedure generates infinite cascades of exactly
solvable $\mathbb{Z}_{2}$ topological dimer liquids beyond the standard
planar framework. The checkerboard model is special in this regard:
since the square lattice is self-medial, the corresponding cascade is
stationary and reproduces the same checkerboard geometry at every step.

While the commuting-projector QDM Hamiltonians defined here represent
special points in their respective phase diagrams, we have also
introduced a controlled way of perturbing such models at the wave
function level by extending Kasteleyn methods to the relevant
non-planar graphs. These wave-function deformations still allow
efficient computation of correlation functions and, in particular, can
interpolate between the commuting-projector ground states and RK states
on planar versions of the lattice. As expected, the Bloch determinant is
exactly flat only at the commuting-projector points and becomes
dispersive in the more generic case. We demonstrated this formalism in
the appendices for the crossed ruby and checkerboard models. In
particular, we verified that it works for the non-planar lattices by
reproducing dimer-counting formulas that can also be obtained directly
from the arrow representation.

More broadly, our results show that fully solvable quantum dimer models
with $\mathbb{Z}_{2}$ topological order and dual gauge-theory
descriptions can persist even on non-planar lattices, provided suitable
local constraints are imposed. This considerably enlarges the class of geometries on  which analytically controlled topological dimer liquids can be constructed. Because the Hilbert spaces of these models remain relatively moderate in size, they should also provide attractive platforms for future numerical work, including studies of perturbations away from the exactly solvable point, nonequilibrium dynamics, and possible quantum many-body scar phenomena in constrained topological systems~\cite{wildeboer2020topological}.

\section{Acknowledgements}
JW was supported by Office of Basic Energy Sciences, Material Sciences and Engineering Division, U.S. Department of Energy (DOE) under Contracts No. DE-SC0012704. 
JW acknowledges the hospitality of the Perimeter Institute for Theoretical Physics as a Simons Emmy Noether Fellow, during which part of this work was completed. Research at Perimeter Institute is supported by the Government of Canada through the Department of Innovation, Science and Economic Development Canada and by the Province of Ontario through the Ministry of Colleges and Universities. AS has been supported in part by the National Science Foundation under Grant No.\ DMR-2029401, and by the  Ralph W. \& Margery S. Hufferd Gift.





\appendix
\section{Kasteleyn formulation for ruby lattices}\label{app:ruby_pfaffian}

\begin{figure}[t]
\includegraphics[width=1.00\columnwidth]{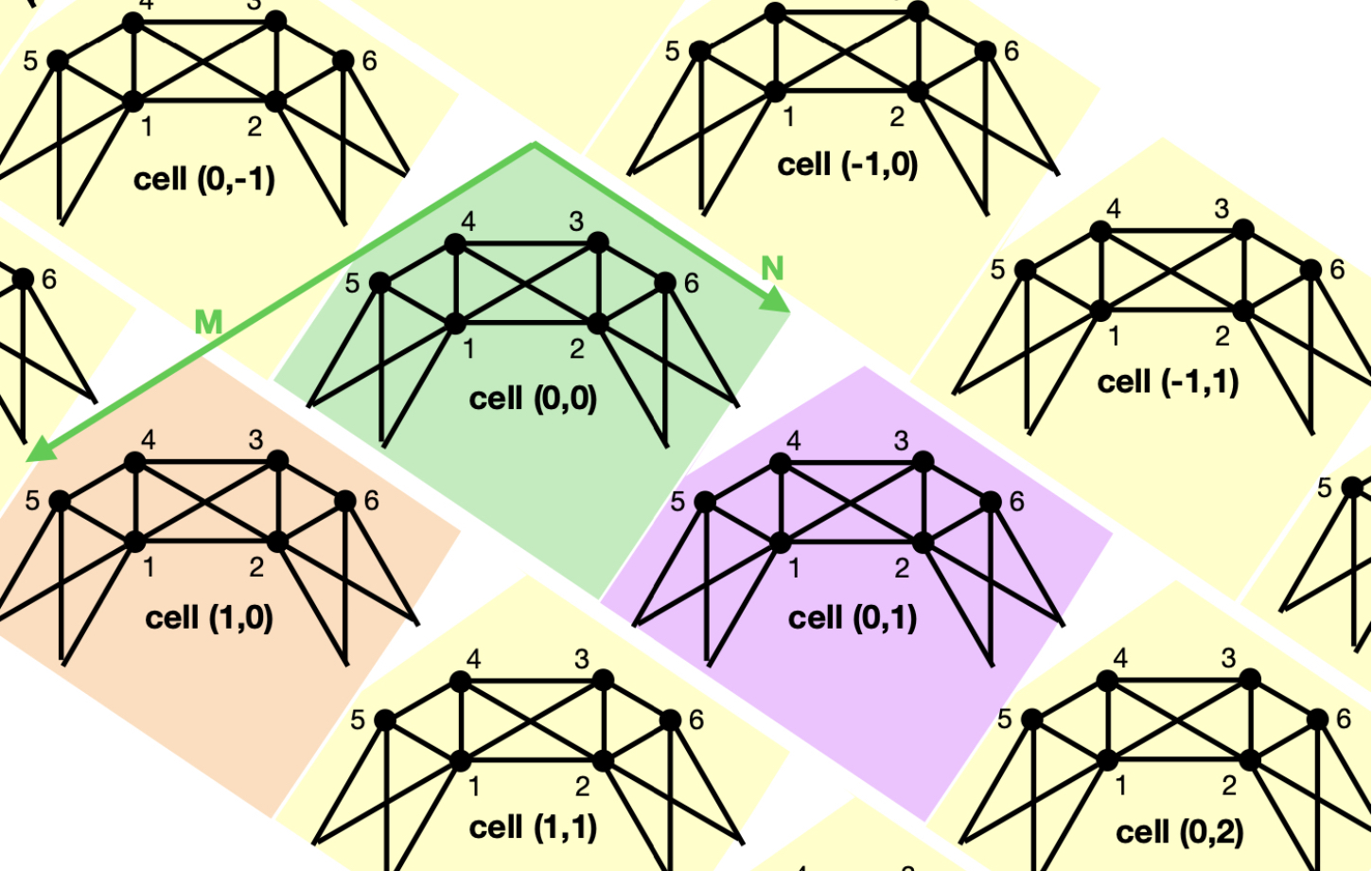}
\caption{ 
         {\it Cell-indexing convention for the ruby-lattice Bloch construction.} 
         Tiling of the crossed ruby lattice by the six-site primitive unit cell, showing the cell-indexing convention and the lattice directions used in the Bloch construction. The central cell is labeled $(0,0)$, and the neighboring cells relevant for the primitive blocks are shown explicitly. The translation directions associated with the Bloch angles $\phi$ and $\theta$ are denoted by $M$ and $N$, respectively. This figure fixes the real-space convention used in Appendix~\ref{app:ruby_pfaffian} for the construction of the matrices $a_{00}$, $a_{10}$, and $a_{01}$, and hence of the primitive Bloch kernels for the regular and crossed ruby lattices. 
}\label{fig:ruby8d}
\end{figure}

In this Appendix we derive expressions for the number of close-packed hard-core dimer matchings (``dimer coverings'') on ruby lattices with periodic boundary conditions. Appendix~\ref{app:regular_ruby} treats the regular ruby lattice (planar, i.e. without diagonals in the square/rectangular plaquettes) using a standard Kasteleyn orientation as shown in Fig.~\ref{fig:ruby8abc}(b) and the Pfaffian method. Appendix~\ref{app:spec_6site} treats the crossed ruby lattice with diagonals and the constraint that parallel two-dimer occupancies on a crossed rectangle are forbidden, which requires a pre-Kasteleyn setup [see Fig.~\ref{fig:ruby8abc}(c)]. Throughout, we work with an $M\times N$ torus of primitive unit cells as depicted in Fig.~\ref{fig:ruby8d} and construct the four sectors corresponding to periodic/antiperiodic twists around the two torus cycles.

\subsection{The regular ruby lattice}\label{app:regular_ruby}

The close-packed hard-core dimer model on any planar lattice can be solved using Kasteleyn--Pfaffian techniques. The key is to orient the edges so that every even-length elementary face is clockwise odd, i.e.  when traversing the boundary of the face clockwise, an odd number of arrows point against the direction of traversal. With such an orientation one builds an antisymmetric Kasteleyn matrix $A$, and the dimer partition function $Z$, i.e. the number of perfect matchings, is given by the Pfaffian of $A$.

For open boundary conditions (OBC), one has
\begin{equation}
Z=\pm {\rm Pf}(A)\,.
\label{eq:A_reg_open}
\end{equation}
On a torus with periodic boundary conditions (PBC), four inequivalent twist sectors appear. Equivalently, one reverses the arrows on a pair of non-contractible boundary ribbons along the two torus cycles, leading to four antisymmetric Kasteleyn matrices $A_{\mu\nu}$ with
$(\mu,\nu)\in\{0,1\}^{2}$. The torus partition function is then assembled from the four Pfaffians in the standard way, i.e. 
\begin{align}
\begin{split}
&Z_{\mathrm{torus}}(M,N) = \\
&\qquad\quad\frac12\Big[
-{\rm Pf}(A_{00}) + {\rm Pf}(A_{01}) + {\rm Pf}(A_{10}) + {\rm Pf}(A_{11})
\Big]\,.
\label{eq:A_reg_torus_pf}
\end{split}
\end{align}
Since each sector Pfaffian satisfies
\begin{equation}
{\rm Pf}(A_{\mu\nu})=\pm \sqrt{\det A_{\mu\nu}}\,,
\label{eq:A_reg_pf_sqrt_det}
\end{equation}
one may equivalently write
\begin{equation}
Z_{\mathrm{torus}}(M,N) =
\frac12\sum_{\mu,\nu\in\{0,1\}}
\sigma_{\mu\nu}\sqrt{\det A_{\mu\nu}} 
\label{eq:A_reg_torus_det}
\end{equation}
with $\sigma_{\mu\nu}\in\{\pm1\}$. The signs $\sigma_{\mu\nu}$ are fixed by direct small-size enumeration in the present ribbon/orientation convention. 

We work on the regular ruby lattice, i.e. the planar ruby lattice without diagonal bonds in the rectangular plaquettes, see Fig.~\ref{fig:ruby7}(a). We use the same six-site primitive cell and the same lattice-direction convention as in the special-ruby case, with $\phi$ associated with translation in the $M$-direction and $\theta$ associated with translation in the $N$-direction. The primitive cell is the six-site bridge shown in Fig.~\ref{fig:ruby8abc}(b), with sites labeled $1,\dots,6$. Bond weights are denoted by $x$ on the long rectangular bonds and by $y$ on the short triangle bonds. The regular-ruby orientation is obtained from the crossed-ruby orientation by removing the diagonal $z$-bonds. Since the resulting graph is planar, this becomes an ordinary Kasteleyn orientation.

Given an oriented weighted edge $ij$ with weight
$w_{ij} \in \{x,y\}$, we use the convention
\begin{equation}
A_{ij}=
\begin{cases}
+w_{ij}, & \text{if the oriented bond points } i\to j,\\
-w_{ij}, & \text{if it points } j\to i,\\
0, & \text{otherwise},
\end{cases}
\label{eq:A_reg_entry_rule}
\end{equation}
so that $A^T=-A$. 

With a six-site unit cell, the total number of sites is $6MN$ for an $M\times N$ torus of unit cells. Using the standard block-Toeplitz construction, the four $6MN\times 6MN$ Kasteleyn matrices can be written as 
\begin{align}
\begin{split}
A_{00} &= a_{00}\otimes I_M\otimes I_N\\
&+a_{10}\otimes T_M\otimes I_N
-a_{10}^{T}\otimes T_M^{T}\otimes I_N \\
&+a_{01}\otimes I_M\otimes T_N
-a_{01}^{T}\otimes I_M\otimes T_N^{T},
\\[1mm]
A_{01} &= a_{00}\otimes I_M\otimes I_N\\
&+a_{10}\otimes H_M\otimes I_N
-a_{10}^{T}\otimes H_M^{T}\otimes I_N \\
&+a_{01}\otimes I_M\otimes T_N
-a_{01}^{T}\otimes I_M\otimes T_N^{T},
\\[1mm]
A_{10} &= a_{00}\otimes I_M\otimes I_N\\
&+a_{10}\otimes T_M\otimes I_N
-a_{10}^{T}\otimes T_M^{T}\otimes I_N \\
&+a_{01}\otimes I_M\otimes H_N
-a_{01}^{T}\otimes I_M\otimes H_N^{T},
\\[1mm]
A_{11} &= a_{00}\otimes I_M\otimes I_N\\
&+a_{10}\otimes H_M\otimes I_N
-a_{10}^{T}\otimes H_M^{T}\otimes I_N \\
&+a_{01}\otimes I_M\otimes H_N
-a_{01}^{T}\otimes I_M\otimes H_N^{T}\,.
\end{split}
\label{eq:A_reg_A11}
\end{align}
Here $\otimes$ denotes the tensor product, $I_{M}$ is the $M\times M$  identity matrix, and $T_{N}$ and $H_{N}$ are the $N\times N$ shift matrices implementing periodic and antiperiodic boundary conditions, respectively,
\begin{equation}
H_{N} =
\begin{pmatrix}
0&1&0&\cdots&0\\
0&0&1&\cdots&0\\
\vdots&\vdots&\vdots&\ddots&\vdots\\
0&0&0&\cdots&1\\
-1&0&0&\cdots&0
\end{pmatrix},
\qquad
T_{N} =
\begin{pmatrix}
0&1&0&\cdots&0\\
0&0&1&\cdots&0\\
\vdots&\vdots&\vdots&\ddots&\vdots\\
0&0&0&\cdots&1\\
1&0&0&\cdots&0
\end{pmatrix}\,.
\label{eq:A_reg_shift_mats}
\end{equation}
Analogous definitions hold for matrices $T_{M}$ and $H_{M}$. 

All bonds within the unit cell and those connecting to
neighboring cells are encoded in the $6\times6$ matrices 
\begin{equation}
a_{00}=
\begin{pmatrix}
0 & x & 0 & -y & y & 0\\
-x & 0 & y & 0 & 0 & -y\\
0 & -y & 0 & -x & 0 & y\\
y & 0 & x & 0 & -y & 0\\
-y & 0 & 0 & y & 0 & 0\\
0 & y & -y & 0 & 0 & 0
\end{pmatrix},
\label{eq:A_reg_a00}
\end{equation}
and 
\begin{align}
\begin{split}
a_{01}=
\begin{pmatrix}
0&0&0&0&0&0\\
0&0&0&0&x&0\\
0&0&0&0&0&0\\
0&0&0&0&0&0\\
0&0&0&0&0&0\\
0&0&0&x&0&0
\end{pmatrix},\quad
a_{10}=
\begin{pmatrix}
0&0&0&0&0&x\\
0&0&0&0&0&0\\
0&0&0&0&0&0\\
0&0&0&0&0&0\\
0&0&x&0&0&0\\
0&0&0&0&0&0
\end{pmatrix},
\label{eq:A_reg_a01_a10}
\end{split}
\end{align}
together with
\begin{equation}
a_{0,-1}=-a_{01}^{T},
\qquad
a_{-1,0}=-a_{10}^{T}\,.
\label{eq:A_reg_backward_blocks}
\end{equation}

To evaluate the four sector determinants, we Fourier transform
the block-Toeplitz structure. 
For periodic boundary conditions ($\mu=\nu=0$), the shift operators have eigenvalues given by
\begin{align} 
\begin{split} 
&T_N \mapsto e^{i\theta_{n,\mu=0}} \quad \text{and} \quad T_N^\dagger \mapsto e^{-i\theta_{n,\mu=0}}, \\ 
&T_M \mapsto e^{i\phi_{m,\nu=0}} \quad \text{and} \quad T_M^\dagger \mapsto e^{-i\phi_{m,\nu=0}},
\end{split} 
\label{eq:A_reg_periodic_eigs} 
\end{align}
whereas for antiperiodic boundary conditions ($\mu=\nu=1$), the corresponding eigenvalues are
\begin{align} 
\begin{split} 
&H_{N} \mapsto e^{i\theta_{n,\mu=1}} \quad \text{and} \quad H_{N}^{\dagger} \mapsto e^{-i\theta_{n,\mu=1}}, \\ 
&H_{M} \mapsto e^{i\phi_{m,\nu=1}} \quad \text{and} \quad H_{M}^{\dagger} \mapsto e^{-i\phi_{m,\nu=1}}\,.
\end{split} 
\label{eq:A_reg_antiperiodic_eigs} 
\end{align}
Here, the continuous spectra are parameterized by the grids
\begin{align} 
\begin{split} 
&\theta_{n,\mu}=\frac{(2n+\mu)\pi}{N}, \quad n=0,\ldots,N-1 \\ 
&\phi_{m,\nu}=\frac{(2m+\nu)\pi}{M}, \quad m=0,\ldots,M-1
\end{split} 
\label{eq:A_reg_theta_phi_grids} 
\end{align}
with  $(\mu,\nu)\in\{0,1\}^{2}$. 
\begin{table}[b]
\centering
\begin{ruledtabular}
\begin{tabular}{c c c r @{\hspace{1cm}} c c c r}
$M$ & $N$ & $6MN$ & $Z_{\mathrm{torus}}$ & $M$ & $N$ & $6MN$ & $Z_{\mathrm{torus}}$ \\
1 & 1 & 6   & 8               & 3 & 6 & 108 & 1338883969024 \\
2 & 1 & 12  & 40              & 4 & 2 & 48  & 376960        \\
2 & 2 & 24  & 896             & 4 & 3 & 72  & 152759936     \\
3 & 2 & 36  & 18016           & 4 & 4 & 96  & 64346914816   \\
3 & 3 & 54  & 1640576         & 5 & 2 & 60  & 7920000       \\
3 & 4 & 72  & 152759936       & 5 & 3 & 90  & 14290363776   \\
3 & 5 & 90  & 14290363776     & 6 & 2 & 72  & 167321600     \\
\end{tabular}
\end{ruledtabular}
\caption{
        {\it Exact torus counts for the regular ruby lattice.} 
        Total number of dimer coverings $Z_{\mathrm{torus}}$ on the regular ruby lattice with periodic boundary conditions, for tori consisting of $M\times N$ primitive unit cells and hence $6MN$ lattice sites. The entries are evaluated at isotropic link weights $x=y=1$ using the four-sector determinant formula $Z_{\mathrm{torus}}(M,N)=\frac12\sum_{\mu,\nu\in\{0,1\}}\sqrt{\det A_{\mu\nu}}$. In contrast to the crossed ruby case, these values arise from a genuinely dispersive Bloch determinant and nontrivial Brillouin-zone products.
}
\label{tab:ruby_regular_total666}
\end{table}
Consequently, the block-Toeplitz determinants factorize into a 
product of $6\times6$ Bloch determinants, 
\begin{equation}
\det A_{\mu\nu} = \prod_{n=0}^{N-1}\prod_{m=0}^{M-1}
\det A(\theta_{n,\mu},\phi_{m,\nu})\,,
\label{eq:A_reg_sector_det}
\end{equation}
where the primitive Bloch matrix is
\begin{align}
\begin{split}
A(\theta,\phi)
&=\sum_{(M_1,N_1)}
a_{M_1,N_1}\,e^{\,i(M_1\phi+N_1\theta)}\\[2pt]
&=
a_{00}
+a_{01}e^{i\theta} -a_{01}^{T}e^{-i\theta}
+a_{10}e^{i\phi} -a_{10}^{T}e^{-i\phi} 
\label{eq:A_reg_bloch_def}
\end{split}
\end{align}
with $(M_1,N_1)\in\{(0,0),(0,1),(1,0)\}$. 
Multiplying out Eq.~\eqref{eq:A_reg_bloch_def} gives
\begin{align}
\begin{split}
&A(\theta,\phi)= \\
&\qquad\begin{pmatrix}
0 & x & 0 & -y & y & x e^{i\phi}\\
-x & 0 & y & 0 & x e^{i\theta} & -y\\
0 & -y & 0 & -x & -x e^{-i\phi} & y\\
y & 0 & x & 0 & -y & -x e^{-i\theta}\\
-y & -x e^{-i\theta} & x e^{i\phi} & y & 0 & 0\\
-x e^{-i\phi} & y & -y & x e^{i\theta} & 0 & 0
\end{pmatrix}\,.
\label{eq:A_reg_bloch_explicit}
\end{split}
\end{align}

As in the crossed-ruby case, the Bloch block is not generically skew-symmetric at fixed $(\theta,\phi)$. Rather, one has
\begin{equation}
A(\theta,\phi)^T=-A(-\theta,-\phi) 
\label{eq:A_reg_bloch_skew}
\end{equation}
so the natural primitive per-mode object is $\det A(\theta,\phi)$, while Pfaffians are taken only after assembling the full real-space matrices $A_{\mu\nu}$. 

For real $x,y$, the Bloch determinant is real and even. Using Eq.~\eqref{eq:A_reg_bloch_skew}, one has
$\det A(\theta,\phi)=\det A(-\theta,-\phi)$, while complex conjugation sends $e^{i(\cdots)}\mapsto e^{-i(\cdots)}$, so
that $\det A(\theta,\phi)^{*}=\det A(-\theta,-\phi)$ and hence $\det A(\theta,\phi)\in\mathbb{R}$. A direct symbolic evaluation yields the trigonometric polynomial 
\begin{widetext}
\begin{align}
\det A(\theta,\phi)
={}&
2x^2\Big[
2x^4
-2x^2y^2\cos\phi
-x^2y^2\cos(2\phi)
+2x^2y^2\cos\theta
-x^2y^2\cos(2\theta)
\nonumber\\
&\qquad
+2x^2y^2\cos(\phi-\theta)
+2x^2y^2\cos(\phi+\theta)
-2x^2y^2\cos(\phi+2\theta)
+2x^2y^2\cos(2\phi+\theta)
\nonumber\\
&\qquad
-x^2y^2\cos(2\phi+2\theta)
+3x^2y^2
+4y^4\cos\phi
-4y^4\cos\theta
-4y^4\cos(\phi+\theta)
+6y^4
\Big]\,.
\label{eq:A_reg_det_general}
\end{align}
\end{widetext}

In the isotropic case $x=y=t$, this simplifies to
\begin{widetext}
\begin{align}
\begin{split}
\det A(\theta,\phi)
&=
2t^6\Big[
11
+2\cos\phi
-\cos(2\phi)
-2\cos\theta -\cos(2\theta)
+2\cos(\phi-\theta)
-2\cos(\phi+\theta)\\
&
\quad\quad-2\cos(\phi+2\theta)
+2\cos(2\phi+\theta) -\cos(2\phi+2\theta)
\Big]\,.
\label{eq:A_reg_det_iso}
\end{split}
\end{align}
\end{widetext}
The torus partition sum is obtained from Eq.~\eqref{eq:A_reg_torus_det}, where the coefficients $\sigma_{\mu\nu}$ specify the sign choice relating the sector Pfaffians to the positive square roots of the sector determinants. For the ribbon/orientation convention adopted here, these signs are fixed by explicit enumeration on the smallest tori and are found to be
\begin{equation}
\sigma_{\mu\nu}=+1
\qquad
\forall\,(\mu,\nu)\in\{0,1\}^{2}\,.
\label{eq:A_reg_allplus}
\end{equation}
Hence
\begin{equation}
Z_{\mathrm{torus}}(M,N)=
\frac12\sum_{\mu,\nu\in\{0,1\}}\sqrt{\det A_{\mu\nu}}
\label{eq:A_reg_torus_final}
\end{equation}
since we select the all-plus choice. 

As a concrete illustration of the regular-ruby counting problem, 
Table~\ref{tab:ruby_regular_total666} lists the resulting torus partition sums $Z_{\mathrm{torus}}$ for several finite system sizes at isotropic link weights $x=y=1$. In contrast to the crossed-ruby case discussed below, these values arise from a genuinely dispersive primitive Bloch determinant and therefore from nontrivial Brillouin-zone products over the four Pfaffian sectors.



\subsection{Crossed ruby lattice: $\mathbf{6}$-site primitive unit cell}
\label{app:spec_6site}

In this subsection we analyze the crossed ruby lattice using the 
six-site primitive unit cell shown in Figs.~\ref{fig:ruby8abc}(c) and ~\ref{fig:ruby8d}. The lattice directions are denoted by $M$ and $N$, and we choose the corresponding Bloch angles so that $\phi$ is associated with translation in the $M$-direction and $\theta$ with translation in the $N$-direction. With this convention, the nontrivial inter-cell couplings occur only in the $(1,0)$ and $(0,1)$ directions, and the natural primitive blocks are therefore $a_{00}$, $a_{10}$, and $a_{01}$. 

As in the main text, the bond weights are denoted by $x$, $y$, and $z$:
$x$ on the long rectangular bonds, $y$ on the short triangle bonds, and
$z$ on the diagonals inside the crossed rectangular plaquettes. We first
write down the primitive blocks and the corresponding $6\times6$ Bloch
kernel explicitly. These expressions form the starting point for the
determinant manipulations in the subsequent parts of the appendix.

\subsubsection{Primitive $6$-site blocks and Bloch Kasteleyn kernel}\label{app:spec_6site_a}

We use the six-site unit cell shown in Figs.~\ref{fig:ruby8abc}(c) and ~\ref{fig:ruby8d}, with site labels $(1,2,3,4,5,6)$ inside each primitive cell. In this basis, the intra-cell block is
\begin{widetext}
\begin{equation}
a_{00}=
\begin{pmatrix}
0 & x & z & -y & y & 0\\
-x & 0 & y & -z & 0 & -y\\
-z & -y & 0 & -x & 0 & y\\
y & z & x & 0 & -y & 0\\
-y & 0 & 0 & y & 0 & 0\\
0 & y & -y & 0 & 0 & 0
\end{pmatrix}\,.
\label{eq:A_spec_a00}
\end{equation}  
\end{widetext}

The inter-cell couplings occur only in the $N$-direction $(0,1)$ and in
the $M$-direction $(1,0)$. The corresponding forward blocks are
\begin{align}
\begin{split}
a_{01}=
\begin{pmatrix}
0&0&0&0&0&0\\
0&0&0&z&x&0\\
0&0&0&0&0&0\\
0&0&0&0&0&0\\
0&0&0&0&0&0\\
0&0&0&x&z&0
\end{pmatrix},
\qquad 
a_{10}=
\begin{pmatrix}
0&0&z&0&0&x\\
0&0&0&0&0&0\\
0&0&0&0&0&0\\
0&0&0&0&0&0\\
0&0&x&0&0&z\\
0&0&0&0&0&0
\end{pmatrix}\,.
\label{eq:A_spec_a01_a10}
\end{split}
\end{align}
Antisymmetry fixes the backward blocks as
\begin{equation}
a_{0,-1}=-a_{01}^{T},
\qquad
a_{-1,0}=-a_{10}^{T}\,.
\label{eq:A_spec_backward_blocks}
\end{equation}

With our convention that $\theta$ is the Bloch angle in the
$N$-direction and $\phi$ the Bloch angle in the $M$-direction, the
primitive Bloch kernel is
\begin{align}
\begin{split}
A_6(\theta,\phi)
&=\sum_{(M_1,N_1)}
a_{M_1,N_1}\,e^{\,i(M_1\phi+N_1\theta)}\\[2pt]
&=
a_{00}
+a_{01}e^{i\theta}-a_{01}^{T}e^{-i\theta}
+a_{10}e^{i\phi}-a_{10}^{T}e^{-i\phi},
\label{eq:A_spec_A6_def}
\end{split}
\end{align}
with $(M_1,N_1)\in\{(0,0),(0,1),(1,0)\}$. Thus the
$a_{01}$ block carries the $N$-direction phase $e^{i\theta}$,
whereas the $a_{10}$ block carries the $M$-direction phase
$e^{i\phi}$.

Expanding Eq.~\eqref{eq:A_spec_A6_def} gives
\begin{widetext}
\begin{equation}
A_{6}(\theta,\phi)=
\begin{pmatrix}
0 & x & z\bigl(1+e^{i\phi}\bigr) & -y & y & x e^{i\phi} \\[1mm]
-x & 0 & y & -z+z e^{i\theta} & x e^{i\theta} & -y \\[1mm]
-z\bigl(1+e^{-i\phi}\bigr) & -y & 0 & -x & - x e^{-i\phi} & y \\[1mm]
y & z-z e^{-i\theta} & x & 0 & -y & - x e^{-i\theta} \\[1mm]
-y & - x e^{-i\theta} & x e^{i\phi} & y & 0 & z e^{i\phi}-z e^{-i\theta}\\[1mm]
- x e^{-i\phi} & y & -y & x e^{i\theta} & - z e^{-i\phi}+z e^{i\theta} & 0
\end{pmatrix}.
\label{eq:A_spec_A6_explicit}
\end{equation}
\end{widetext}

For later manipulations it is convenient to introduce the shorthand
\begin{equation}
\alpha=e^{i\theta},
\qquad
\beta=e^{i\phi},
\label{eq:A_spec_alpha_beta}
\end{equation}
together with their inverses $\alpha^{-1} = e^{-i\theta}$ and $\beta^{-1} = e^{-i\phi}$, so that
Eq.~\eqref{eq:A_spec_A6_explicit} may also be written as
\begin{widetext}
\begin{equation}
A_{6}(\theta,\phi)=
\begin{pmatrix}
0 &
x &
z(1+\beta) &
-y &
y &
x\beta
\\[1mm]
-x &
0 &
y &
-z+z\alpha &
x\alpha &
-y
\\[1mm]
-z(1+\beta^{-1}) &
-y &
0 &
-x &
- x\beta^{-1} &
y
\\[1mm]
y &
z-z\alpha^{-1} &
x &
0 &
-y &
- x\alpha^{-1}
\\[1mm]
-y &
- x\alpha^{-1} &
x\beta &
y &
0 &
z\beta-z\alpha^{-1}
\\[1mm]
- x\beta^{-1} &
y &
-y &
x\alpha &
- z\beta^{-1}+z\alpha &
0
\end{pmatrix}.
\label{eq:A_spec_A6_explicit_ab}
\end{equation}
\end{widetext}

The primitive Bloch kernel is not skew-symmetric at fixed
$(\theta,\phi)$. Rather, it satisfies the Bloch-skew relation
\begin{equation}
A_{6}(\theta,\phi)^T=-A_6(-\theta,-\phi)\,.
\label{eq:A_spec_bloch_skew}
\end{equation}
Accordingly, the natural primitive quantity to study is
$\det A_{6}(\theta,\phi)$ rather than a mode-wise Pfaffian.

\subsubsection{Schur-complement reduction of $\det A_{6}(\theta,\phi)$ for isotropic link weights $t=x=y=z$}\label{app:spec_6site_A33}

We now specialize to the isotropic point $t=x=y=z$ 
and evaluate the determinant of the primitive Bloch kernel~\eqref{eq:A_spec_A6_explicit_ab} by a Schur-complement reduction.

We split the matrix~\eqref{eq:A_spec_A6_explicit_ab} into the central
four-site block $(1,2,3,4)$ and the two side sites $(5,6)$, i.e. 
\begin{equation}
A_{6}(\theta,\phi)
=
\begin{pmatrix}
P(\theta,\phi) & T(\theta,\phi)\\[1mm]
U(\theta,\phi) & S(\theta,\phi)
\end{pmatrix}\,.
\label{eq:A_spec_block_split}
\end{equation}
The $4\times4$ block is
\begin{equation}
P(\theta,\phi)=
\begin{pmatrix}
0 & t & t(1+\beta) & -t\\
-t & 0 & t & t(\alpha-1)\\
-t(1+\beta^{-1}) & -t & 0 & -t\\
t & t-t\alpha^{-1} & t & 0
\end{pmatrix},
\label{eq:A_spec_P}
\end{equation}
while the $4\times2$ block is
\begin{equation}
T(\theta,\phi)=
\begin{pmatrix}
t & t\beta\\
t\alpha & -t\\
-t\beta^{-1} & t\\
-t & -t\alpha^{-1}
\end{pmatrix}\,.
\label{eq:A_spec_T}
\end{equation}
Moreover, the $2\times4$ block is
\begin{equation}
U(\theta,\phi)=
\begin{pmatrix}
-t & -t\alpha^{-1} & t\beta & t\\
-t\beta^{-1} & t & -t & t\alpha
\end{pmatrix}
\label{eq:A_spec_U}
\end{equation}
and the $2\times2$ block is
\begin{equation}
S(\theta,\phi)=
\begin{pmatrix}
0 & t(\beta-\alpha^{-1})\\
t(\alpha-\beta^{-1}) & 0
\end{pmatrix}\,.
\label{eq:A_spec_S}
\end{equation}

For $\alpha\beta \neq 1$, the block $S$ is invertible, with
\begin{equation}
S(\theta,\phi)^{-1}
=
\begin{pmatrix}
0 & \dfrac{\beta}{t(\alpha\beta-1)}\\[3mm]
\dfrac{\alpha}{t(\alpha\beta-1)} & 0
\end{pmatrix}\,.
\label{eq:A_spec_Sinv}
\end{equation} 
The determinant can therefore be reduced by the Schur identity
\begin{equation}
\det A_{6}(\theta,\phi)
=
\det S(\theta,\phi)\cdot
\det P_{\mathrm{eff}}(\theta,\phi) 
\label{eq:Schur_def_Peff}
\end{equation}
with the effective $4\times4$ block
\begin{equation}
P_{\mathrm{eff}}(\theta,\phi)
\equiv
P(\theta,\phi)-T(\theta,\phi)\,S(\theta,\phi)^{-1}\,U(\theta,\phi)\,.
\label{eq:Schur_def_Peff666}
\end{equation}
A direct calculation gives
\begin{align}
\begin{split}
&P_{\mathrm{eff}}(\theta,\phi)\\
&\,=
\begin{pmatrix}
\dfrac{t(\alpha\beta+1)}{\alpha\beta-1} & t & t &
\dfrac{t(1-3\alpha\beta)}{\alpha\beta-1}
\\[3mm]
-t &
-\dfrac{t(\alpha\beta+1)}{\alpha\beta-1} &
\dfrac{t(3\alpha\beta-1)}{\alpha\beta-1} &
-t
\\[3mm]
-t &
\dfrac{t(3-\alpha\beta)}{\alpha\beta-1} &
-\dfrac{t(\alpha\beta+1)}{\alpha\beta-1} &
-t
\\[3mm]
\dfrac{t(\alpha\beta-3)}{\alpha\beta-1} &
t &
t &
\dfrac{t(\alpha\beta+1)}{\alpha\beta-1}
\end{pmatrix}.
\label{eq:A_spec_Peff}
\end{split}
\end{align}
The determinant of the $2\times2$ block is
\begin{equation}
\det S(\theta,\phi)
=
-t^2\,\frac{(\alpha\beta-1)^2}{\alpha\beta}
\label{eq:A_spec_detS}
\end{equation}
while the determinant of the Schur complement is
\begin{equation}
\det P_{\mathrm{eff}}(\theta,\phi)
=
-64\,\frac{\alpha\beta\,t^4}{(\alpha\beta-1)^2}\,.
\label{eq:A_spec_detPeff}
\end{equation}
Multiplying the two factors, the $\alpha\beta$-dependence cancels
completely, and one obtains the final result 
\begin{equation}
\det A_6(\theta,\phi)=64\,t^6\,.
\label{eq:A_spec_detA6_constant}
\end{equation}

Equation~\eqref{eq:A_spec_detA6_constant} is therefore the closed form
of the primitive determinant at isotropic link weights. Since both
sides are continuous in $(\theta,\phi)$, the result extends by
continuity to the momenta with $\alpha\beta=1$, where the intermediate
inverse $S(\theta,\phi)^{-1}$ becomes singular.

\subsubsection{Closed form for $\det A_{6}(\theta,\phi)$ at the isotropic point $t=x=y=z$}\label{app:spec_6site_A34}

The Schur-complement reduction derived in the previous 
subsubsection implies that, at the isotropic point $t=x=y=z$, 
the determinant of the primitive Bloch kernel takes the simple 
closed form
\begin{equation}
\det A_{6}(\theta,\phi)=64\,t^6\,.
\label{eq:A_spec_detA6_flat}
\end{equation}
Thus the primitive $6\times6$ determinant is completely flat in
momentum space, i.e. it is independent of both Bloch angles
$\theta$ and $\phi$.

Equation~\eqref{eq:A_spec_detA6_flat} is the central
primitive-cell result for the crossed ruby lattice at isotropic link weights. In particular, it shows that the momentum dependence present in the individual matrix elements of $A_{6}(\theta,\phi)$ cancels exactly in the determinant.

It is important to distinguish this primitive-kernel statement from the full torus counting problem of the constrained crossed-ruby model. The latter requires the assembly of the appropriate sector contributions on the torus and will be discussed in the following subsections.



\subsubsection{Sector products and torus partition sum}\label{app:spec_6site_A35}

We now pass from the primitive Bloch kernel to the torus geometry. For
an $M\times N$ array of primitive cells, the four standard sectors
$(\mu,\nu)\in\{0,1\}^2$ correspond to periodic or antiperiodic boundary
conditions in the $N$- and $M$-directions, respectively. With our
convention that $\theta$ is the Bloch angle in the $N$-direction and
$\phi$ in the $M$-direction, the allowed momenta in sector $(\mu,\nu)$ are
\begin{equation}
\theta_{n,\mu} = \frac{2\pi}{N}\Bigl(n+\frac{\mu}{2}\Bigr)
\label{eq:A_spec_theta_sector}
\end{equation}
with $n=0,1,\dots,N-1$, and
\begin{equation}
\phi_{m,\nu} = \frac{2\pi}{M}\Bigl(m+\frac{\nu}{2}\Bigr)
\label{eq:A_spec_phi_sector}
\end{equation}
with $m=0,1,\dots,M-1$. The sector determinant is then obtained in the
usual way by taking the product of the primitive determinants over the
corresponding momentum grid, i.e. 
\begin{equation}
\det A_{\mu\nu}
=
\prod_{n=0}^{N-1}\prod_{m=0}^{M-1}
\det A_{6}\bigl(\theta_{n,\mu},\phi_{m,\nu}\bigr)\,.
\label{eq:A_spec_sector_det}
\end{equation}

At isotropic link weights $t=x=y=z$, the primitive determinant is flat and is given by 
\begin{equation}
\det A_6(\theta,\phi)=64\,t^6
\label{eq:A_spec_flat_det_repeat}
\end{equation}
so that all four sector determinants are equal, i.e. 
\begin{equation}
\det A_{\mu\nu} = (64\,t^6)^{MN} 
\label{eq:A_spec_sector_det_flat}
\end{equation}
for all $(\mu,\nu)\in\{0,1\}^{2}$. Consequently, the Pfaffian in each
sector has magnitude
\begin{equation}
{\rm Pf}(A_{\mu\nu}) = \pm\,(8\,t^3)^{MN}\,.
\label{eq:A_spec_sector_pf_mag}
\end{equation}

The torus partition sum is assembled from the four sectors in the
standard manner, i.e. 
\begin{align}
\begin{split}
&Z_{\mathrm{torus}}(M,N) =\\
&\qquad\quad
\frac12\, \Big[-{\rm Pf}(A_{00}) + {\rm Pf}(A_{01}) + {\rm Pf}(A_{10}) + {\rm Pf}(A_{11}) \Big]\,.
\label{eq:A_spec_torus_pf_sum}
\end{split}
\end{align}
Equivalently, one may write
\begin{equation}
Z_{\mathrm{torus}}(M,N)
=
\frac12
\sum_{\mu,\nu\,\in\,\{0,1\}}
\sigma_{\mu\nu}\sqrt{\det A_{\mu\nu}}\,.
\label{eq:A_spec_torus_det_form}
\end{equation}
The signs $\sigma_{\mu\nu}$ encode the choice of branch relating
the sector Pfaffians to the positive square roots of the sector
determinants.

The relative Pfaffian signs are fixed by direct enumeration
on the minimal torus $M=N=1$. For the crossed ruby
lattice, this enumeration yields four dimer coverings in each
of the four winding sectors:
\begin{equation}
Z_{00}=Z_{01}=Z_{10}=Z_{11}=4\,.
\label{eq:A_spec_winding_histograms}
\end{equation}
With the ribbon/orientation convention used here, this implies 
${\rm Pf}(A_{00})=-\sqrt{\det A_{00}}$, ${\rm Pf}(A_{01})=\sqrt{\det A_{01}}$, ${\rm Pf}(A_{10})=\sqrt{\det A_{10}}$, and ${\rm Pf}(A_{11})=\sqrt{\det A_{11}}$. 
Equivalently, in the determinant representation
\eqref{eq:A_spec_torus_det_form}, one may choose
\begin{equation}
\sigma_{\mu\nu}=+1
\qquad
\forall\,(\mu,\nu)\in\{0,1\}^2 .
\label{eq:A_spec_allplus}
\end{equation}
Hence the torus partition sum reduces to 
\begin{align}
\begin{split}
Z_{\mathrm{torus}}(M,N)
&=
\frac12 \sum_{\mu,\nu\,\in\,\{0,1\}}\sqrt{\det A_{\mu\nu}} \\
&=
2 \times (8t^3)^{MN}.
\label{eq:A_spec_torus_pf_final}
\end{split}
\end{align}

\section{Torus counting and sector decomposition for the checkerboard model}\label{appB}

In this appendix we summarize the exact torus counting of admissible dimer coverings for the checkerboard model. For an $M\times N$ array of unit cells, the checkerboard lattice contains $4MN$ sites, since the primitive unit cell shown in Fig.~\ref{figure_666}(a) consists of four sites. 

For the symmetric choice of link weights, the torus partition sum is
\begin{equation}
Z_{\rm torus}=2\times(4x^2)^{MN}\,.
\label{eq:checkerboard_total_count}
\end{equation}
At unit weights $x=1$, this reduces to
\begin{equation}
Z_{\rm torus}=2\cdot 4^{MN}\,.
\label{eq:checkerboard_total_count_unit}
\end{equation}
Thus the number of admissible dimer coverings grows exponentially with
the number of unit cells.

This result admits a sharper sector-resolved form. In the isotropic case, the primitive Bloch determinant is momentum-independent, so all four torus Pfaffian sectors have the same magnitude, i.e. 
\begin{equation}
\bigl|{\rm Pf}(A_{\mu\nu})\bigr|=(4x^2)^{MN}\,.
\end{equation}
Moreover, in the convention used in the main text and in
Ref.~\cite{wildeboer2020checkerboard}, the Pfaffian signs are 
\begin{align}
\begin{split}
&\qquad{\rm Pf}(A_{00})=-(4x^2)^{MN}\\
{\rm Pf}(A_{01})&={\rm Pf}(A_{10})={\rm Pf}(A_{11})=(4x^2)^{MN}\,.
\end{split}
\end{align}
Using the standard inverse relations between Pfaffian sectors and topological sectors, one obtains
\begin{equation}
Z_{00}=Z_{10}=Z_{01}=Z_{11}=\frac12\,(4x^2)^{MN}\,.
\label{eq:checkerboard_sector_count}
\end{equation} 
At unit weights, this becomes
\begin{equation}
Z_{ab}=\frac12\,4^{MN}
\label{eq:checkerboard_sector_count_unit}
\end{equation}
with $a,b\in\{0,1\}$. 
Hence the four topological sectors contribute equally, and
\begin{equation}
Z_{\rm torus}=\sum_{a,b\,\in\,\{0,1\}}Z_{ab}=2\cdot 4^{MN}\,.
\end{equation}

For comparison, the corresponding Pfaffian construction for the kagome
lattice was given by Wang and Wu in Ref.~\cite{wangwu}. The main
difference is that the kagome unit cell contains six sites, whereas the
checkerboard unit cell contains four.

The simple closed forms~\eqref{eq:checkerboard_total_count_unit} and  \eqref{eq:checkerboard_sector_count_unit} make the checkerboard model especially convenient for analytical and numerical studies of topological dimer liquids.

\bibliography{bibfile2}
\end{document}